\documentclass[final,3p,times,twocolumn]{elsarticle}
\usepackage[T1]{fontenc}
\usepackage[utf8]{inputenc}
\usepackage{amssymb}
\usepackage{amsmath}
\usepackage{placeins}
\journal{Nuclear Instruments and Methods in Physics Research Section A}
\usepackage[hidelinks]{hyperref}
\usepackage{lineno}
\usepackage{url}

\usepackage{xcolor}
\usepackage{ulem}

\begin{document}
% \maketitle
% \flushbottom

\begin{frontmatter}

\title{CTC and CT5TEA: an advanced multi-channel digitizer and trigger ASIC for imaging atmospheric Cherenkov telescopes \footnotemark}%\thanks{This paper is dedicated to the memory of Gary Varner, creator of the TARGET ASIC family}}

\author[a]{B. Schwab\corref{cor1}}
\ead{ben.schwab@fau.de}
\author[a]{A. Zink\corref{cor1}}
\ead{adrian.zink@fau.de}
\cortext[cor1]{Corresponding author}
\author[b]{G. Varner\corref{cor2}}
\cortext[cor2]{Deceased}

\author[c]{D. Depaoli}
\author[c]{J. Hinton}
\author[b,d]{G. Liu}
\author[e]{A. Okumura}
\author[f]{D. Ross} %affiliation unclear
\author[a]{J. Sch\"afer}
\author[g]{H. Schoorlemmer}
\author[e]{H. Tajima}
\author[h,i]{J. Vandenbroucke}
\author[c]{R. White}
\author[j]{J.J. Watson}
\author[c,k]{J. Zorn} %affiliation unclear

% %Stefan as last one
\author[a]{S. Funk}

% The "\note" macro will give a warning: "Ignoring empty anchor..."
% you can safely ignore it.
% \affiliation[a]{Erlangen Centre for Astroparticle Physics (ECAP), Friedrich-Alexander-Universit\"at Erlangen-N\"urnberg, Nikolaus-Fiebiger-Str. 2, 91058 Erlangen, Germany}

\affiliation[a]{organization={Erlangen Centre for Astroparticle Physics (ECAP), Friedrich-Alexander-Universit\"at Erlangen-N\"urnberg},
            addressline={\mbox{Nikolaus-Fiebiger-Str. 2}},
            city={Erlangen},
            postcode={91058},
            country={Germany}}
\affiliation[b]{organization={Department of Physics and Astronomy, University of Hawaii},
            city={Honolulu},
            postcode={96822},
            state={Hawaii},
            country={USA}}
\affiliation[c]{organization={Max Planck Institut f\"ur Kernphysik},
            addressline={\mbox{Saupfercheckweg 1}},
            city={Heidelberg},
            postcode={69117},
            country={Germany}}
\affiliation[d]{organization={Now at  SLAC National Accelerator Laboratory},
            addressline={\mbox{ 2575 Sand Hill Rd Mailstop 0094}},
            city={Menlo Park},
            postcode={94025},
            country={California, USA}}
\affiliation[e]{organization={Institute for Space-Earth Environmental Research, Nagoya University},
            addressline={\mbox{}},
            city={Furo-cho, Chikusa-ku, Nagoya, Aichi},
            postcode={464-8601},
            country={Japan}}
\affiliation[f]{organization={University of Leicester,  Space Research Centre},
            addressline={\mbox{Space Park Leicester LE4 5SP}},
            %city={Zeuthen},
            %postcode={15738},
            country={United Kingdom}}
\affiliation[g]{organization={Institute for Mathematics, Astrophysics and Particle Physics (IMAPP), Radboud University Nijmegen},
            addressline={\mbox{ Houtlaan 4}},
            city={Nijmegen},
            postcode={6525},
            country={Netherlands}}
            %IMAPP, Radboud University Nijmegen, Nijmegen, The Netherlands
\affiliation[h]{organization={Department of Physics, University of Wisconsin-Madison},
            addressline={Madison, WI 53706, USA}}
\affiliation[i]{organization={Wisconsin IceCube Particle Astrophysics Center},
            addressline={ Madison, WI, 53703, USA}}
\affiliation[j]{organization={Deutsches Elektronen-Synchrotron},
            addressline={\mbox{ Platanenallee 6}},
            city={Zeuthen},
            postcode={15738},
            country={Germany}}

\affiliation[k]{organization={Now at Siemens Mobility, Inc},
            addressline={\mbox{Duisburger Straße 145}},
            city={Krefeld},
            postcode={47829},
            country={Germany}}

% e-mail addresses: only for the corresponding author
%\emailAdd{adrian.zink@fau.de, ben.schwab@fau.de}

\begin{abstract}
    We have developed a new set of Application-Specific Integrated Circuits (ASICs) of the TARGET family (CTC and CT5TEA), designed for the readout of signals from photosensors in cameras of Imaging Atmospheric Cherenkov Telescopes (IACTs) for ground-based gamma-ray astronomy. We present the performance and design details. Both ASICs feature 16 channels, with CTC being a Switched-Capacitor Array (SCA) sampler at 0.5 to 1\,GSa/s with a 16,384 sample deep storage buffer, including the functionality to digitize full waveforms at arbitrary times. CT5TEA is its companion trigger ASIC (though may be used on its own), which provides trigger information for the analog sum of four (and 16) adjacent channels. Since sampling and triggering takes place in two separate ASICs, the noise due to interference from the SCA is suppressed, and allows a minimal trigger threshold of $\leq$ 2.5\,mV (0.74\,photo electrons (p.e.)) with a trigger noise of $\leq$ 0.5\,mV (0.15\,p.e.). For CTC, a maximal input voltage range from $-$0.5\,V up to 1.7\,V is achieved with an effective bit range of $>$ 11.6\,bits and a baseline noise of 0.7\,mV. The cross-talk improved to $\leq$ 1\,\% over the whole $-$3\,dB bandwidth of 220\,MHz and even down to 0.2\,\% for 1.5\,V pulses of 10 \,ns width. Not only is the performance presented, but a temperature-stable calibration routine for pulse mode operation is introduced and validated. The resolution is found to be $\sim$ 2.5\,\% at 33.7\,mV (10\,p.e.) and $\leq$ 0.3\% at 337\,mV (100\,p.e.) with an  integrated non-linearity of \mbox{$<$ 1.6\,mV}. Developed for the Small-Sized Telescope (SST) and Schwarzschild-Couder Telescope (SCT) cameras of the Cherenkov Telescope Array Observatory (CTAO), CTC and CT5TEA are deployed for both prototypes and shall be integrated into the final versions.
    %, although baseline noises $<$ 0.6\,mV are achievable for smaller dynamic ranges. 
\end{abstract}


%\keywords{Data acquisition circuits, Trigger concepts and systems (hardware and software), Gamma telescopes}

%\arxivnumber{1234.5678} % only if you have one

%\collaboration[c]{and others TBD}

\end{frontmatter}
\footnotetext{This paper is dedicated to the memory of Gary Varner, creator of the TARGET ASIC family}

\begin{figure*}[t]
\centering
\includegraphics[width=1.0\textwidth]{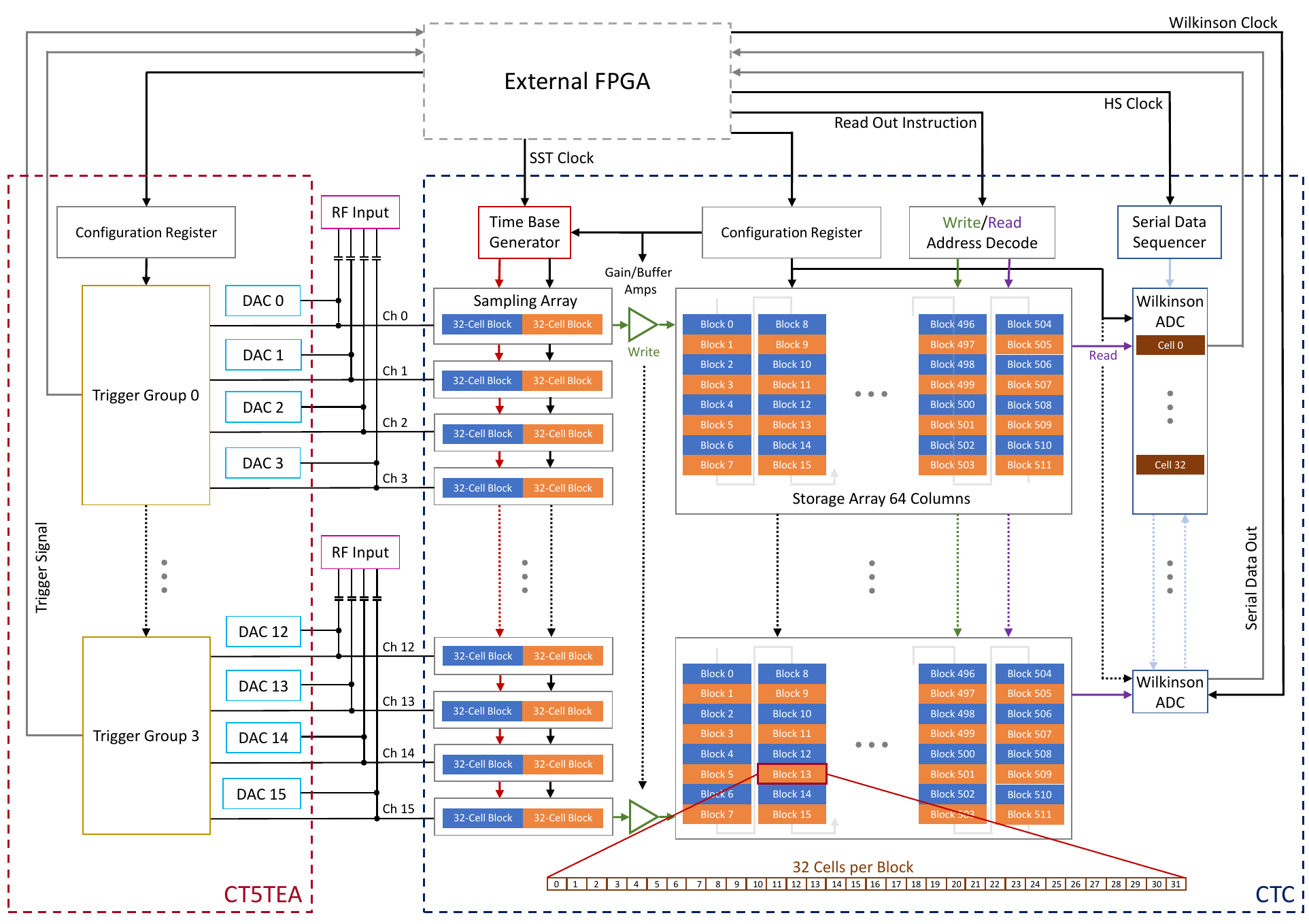}
\caption{\label{fig:ASIC_schematic} Operational schematic of the CT5TEA and CTC ASICs. The analog signal of each channel is fed in parallel into both ASICs. DACs of the CT5TEA create a DC baseline voltage for each channel, the operational pedestal voltage, to set the working point of the CMOS based SCA of CTC. Four channels form a trigger group in CT5TEA, sending a trigger signal to an accompanying external FPGA if the analog sum of these four channels exceeds the trigger threshold. The sampling arrays in CTC sample continuously in ping-pong mode, namely, while one block is sampling, the other transfers the sampled voltages to the storage cell array. The storage blocks can be arbitrarily read out on demand, tapping the cell voltages to the Wilkinson ADCs. The digitized values are transferred to the FPGA via a serial data interface per channel, timed by the HS clock at 93.75\,MHz.}
\end{figure*}

\section{Introduction \label{sec:intro}}

The Cherenkov Telescope Array Observatory (CTAO) will be the next generation Imaging Atmospheric Cherenkov Telescope (IACT) array. It will exceed the sensitivity of current IACTs by an order of magnitude in the energy range of 100\,GeV to 10\,TeV \citep{Science_with_CTA} and will extend the IACT energy coverage down to 20\,GeV and up to 300\,TeV. Three different-sized telescope classes are developed to achieve these goals. Two of them, the Small-Sized Telescopes (SST) \citep{SPIE_sst_paper} and the Schwarzschild-Couder Telescope (SCT) \citep{dualmirror}, a medium sized telescope, deploy a dual-mirror design for better off-axis optical performance and smaller camera size, allowing the use of silicon photomultipliers (SiPM). This reduces the overall cost of the camera while improving the angular resolution, field of view and, therefore, the gamma-hadron separation \citep{WOOD201611}.

The extended energy coverage demands a large dynamic range and good intensity resolution up to multiple thousands of photons. Faint showers require a low trigger threshold of a few ($\sim$ 10) photo electrons (p.e.). The short duration of the Cherenkov showers ($\sim$ 10\,ns) defines the need for a suitable sampling rate ($\leq$ 250\,MSa/s). Due to the high number of electronics channels ($\sim$ 37 SSTs $\cdot$ 2048 pixel), attention must be paid to compactness and cost efficiency.

A solution to process these signals of the SiPMs is presented in the latest generation of TeV Array Readout Electronics with GSa/s sampling and Event Trigger (TARGET) application-specific integrated circuits (ASICs). The development started with TARGET 1, whose performance is characterized in \citep{Bechtol_2012}. Various improvements and design iterations resulted in TARGET 5, which is presented in \citep{TARGET_5_paper}, and ultimately to the ASIC pair Cherenkov TARGET C (CTC), for sampling and digitization, and Cherenkov TARGET 5 Trigger Extension ASIC (CT5TEA), its companion trigger. Together, they form a compact, cost-efficient and reliable option with broad functionality and are currently deployed for the SST camera \citep{depaoli2023status} and the upgraded prototype SCT camera \citep{pSCT}. A single pair of CTC and CT5TEA combined with an FPGA is sufficient to set up a 1\,GSa/s waveform digitizer with self triggering capabilities.

Key features are 16 channels per ASIC, an adjustable sampling rate between 0.5 and 1\,GSa/s, 16,384 sample deep storage resulting in $\approx$ 16\,$\mu$s buffer time for telescope and coincidence trigger between distant telescopes, continuous sampling without dead time, random access full waveform readout, input voltage range of 2.2\,V digitized in 12 bits, mean power consumption of 60\,mW per channel, single-shot trigger with adjustable threshold and output signal width.

The paper is organized as follows. Section \ref{sec:arch} describes the architecture of both ASICs. Section \ref{sec:evaluationboard} provides an overview of the hardware and software used. Section \ref{sec:perf} is split into calibration and performance of the trigger and digitizer path with a comparison to previous generations of TARGET at the end. 
An overview of the different applications of TARGET is given in Section \ref{sec:FEE}. The conclusion and outlook are drawn in Section \ref{sec:conclusion}.

\section{The Cherenkov TARGET C architecture\label{sec:arch}}

The functional blocks of the ASICs are trigger (CT5TEA), analog sampling with a 16,348 deep storage buffer and analog-to-digital converters (ADCs) for signal digitization (CTC) and digital-to-analog converters (DACs) for internal operation settings. The trigger and sampling paths are split into two separate ASICs, reducing cross-talk between the sampling and trigger paths, providing a lower trigger threshold and larger dynamic range \citep{TC_paper, TARGET_proceeding}. Figure \ref{fig:ASIC_schematic} presents an overview of both signal paths.

Each ASIC processes 16 individual channels in parallel. An adjustable DC baseline per channel, the operational pedestal voltage, is set by DACs in CT5TEA. By summing the analog signals of four adjacent channels, four trigger groups are formed per CT5TEA. If the sum in one trigger group exceeds an adjustable threshold, a trigger signal is output as Low-Voltage Differential Signaling (LVDS). An FPGA can then collect these outputs for further processing as for example giving out readout instructions to CTC.

For the sampling in CTC, a small Switched-Capacitor Array (SCA) of 32 + 32 cells is used to keep the capacitive load at a minimum. The cells are following the signal until disconnected sequentially with a 1\,ns delay, retaining a proportional charge. The 1\,ns delay between the sampling cells result in a nominal sampling speed of 1\,GSa/s. The 64 1\,ns delays are generated by a temperature-stabilized internal time base. A delay-locked loop keeps the transition from last to first sampling cell in phase with an external 15.625\,MHz clock (SST clock). The SCA is arranged in two blocks of 32 cells, allowing it to store one block while the other block samples the applied signal to avoid any dead time (ping-pong mode). The readout signal is transferred to the 16,348 sample-deep ring buffer storage array arranged in 512 blocks of 32 cells, buffering 16\,$\mu$s at 1\,GSa/s. The samples are continuously overwritten until a waveform is to be read out. Then, the blocks where the trigger occurred will no longer be overwritten while the sampling continues for the rest of the storage buffer. Blocks of 32 are random-accessible for digitization.

In the case of a readout/digitization instruction, each cell of the 32-cell blocks is digitized in parallel by Wilkinson ADCs. Here, the storage cell voltage is compared with a reference voltage ramp produced by charging an external capacitor with a constant current, which defines the slope of the ramp. At the start of the digitization, a 12-bit Gray code counter is reset and counts up with the speed of the externally applied Wilkinson clock. When the ramp voltage gets larger than the cell voltage, the counter's value is written to a register as the digitized value. With the currently used Wilkinson Clock of 208\,MHz, it takes about 20\,$\mu$s to digitize a full block \citep{TARGET_5_paper}.

\section{Evaluation board and software}
\label{sec:evaluationboard}

\begin{figure}[t]
\centering
\includegraphics[width=0.48\textwidth]{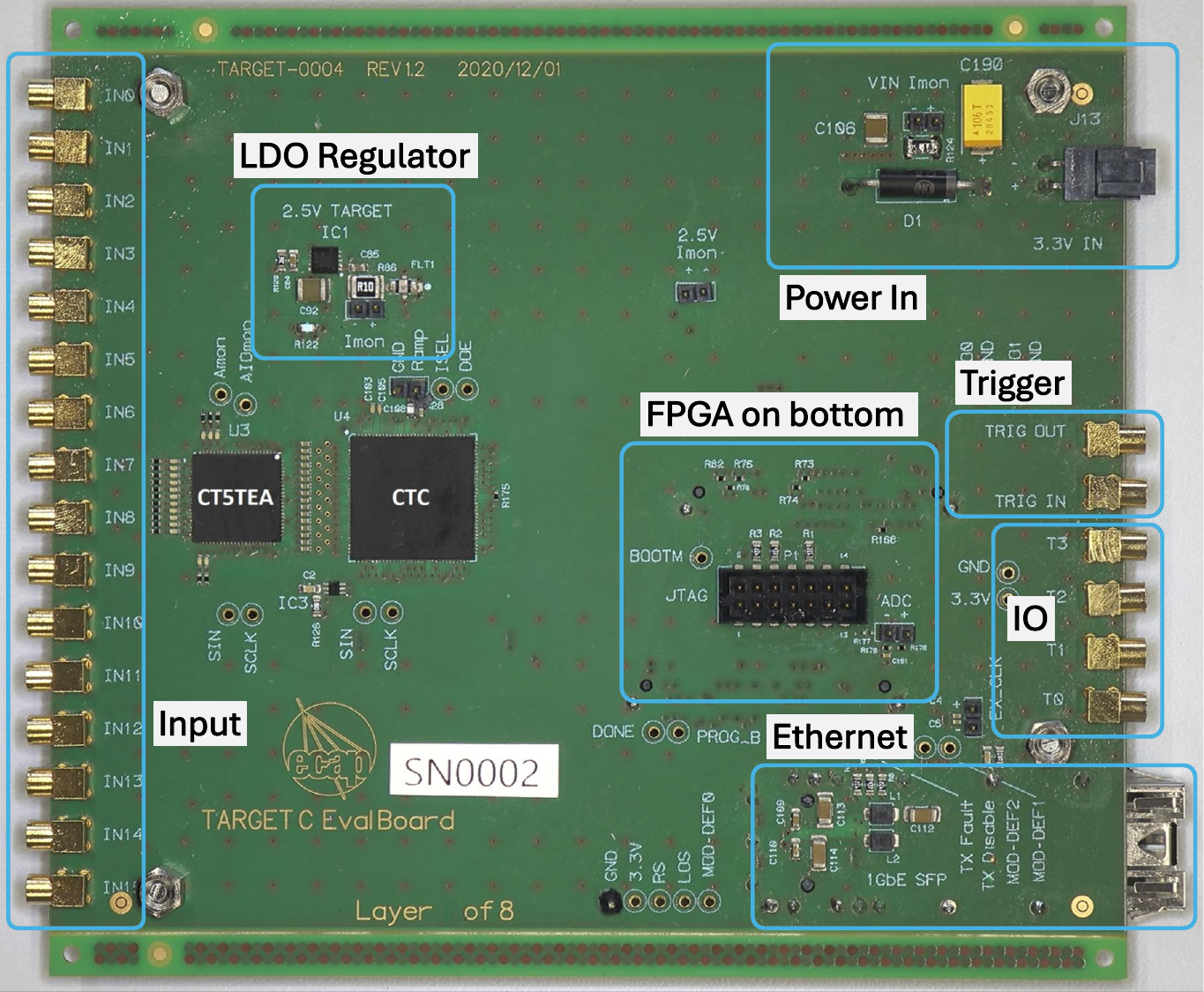}
\caption{\label{fig:evaluation_board} One of the two evaluation boards used to operate CTC and CT5TEA with the FPGA and ethernet connector (SFP port) on the back side, 16 MMCX input connectors on the left, four user-configurable IO ports and Trigger input and output on the right. The board is supplied with a single 3.3V supply (Molex Nano-Fit connector top right).}
\end{figure}

An evaluation board has been developed for the performance evaluation of CTC and CT5TEA, which customizes all elements needed to operate the ASICs. This board is shown in Figure \ref{fig:evaluation_board}. The control of the ASICs is handled by an FPGA piggyback board (Trenz TE0714-03- 35-2I) featuring an Xilinx Artix 7. A gigabit Ethernet link is used for slow control and data transmission to a computer. There is an MMCX connector for each input into the 16 channels of TARGET, terminated with a 50\,$\Omega$ resistor and AC coupled via a 1\,nF capacitor. A low-noise, 2.5\,V output low-dropout (LDO) regulator (Analog Devices LT3045) is placed in close proximity to power the ASICs. The board can be triggered by an external signal, fed from an MMCX connector or created by the FPGA, based on signals from the CT5TEA or the FPGA itself. Additionally, there are four MMCX connectors where signals and functions of the ASICs and FPGA can be monitored. In this work, two evaluation boards were used, designated SN0001 and SN0002.\\

\begin{figure}[t]
\centering
\includegraphics[width=0.48\textwidth]{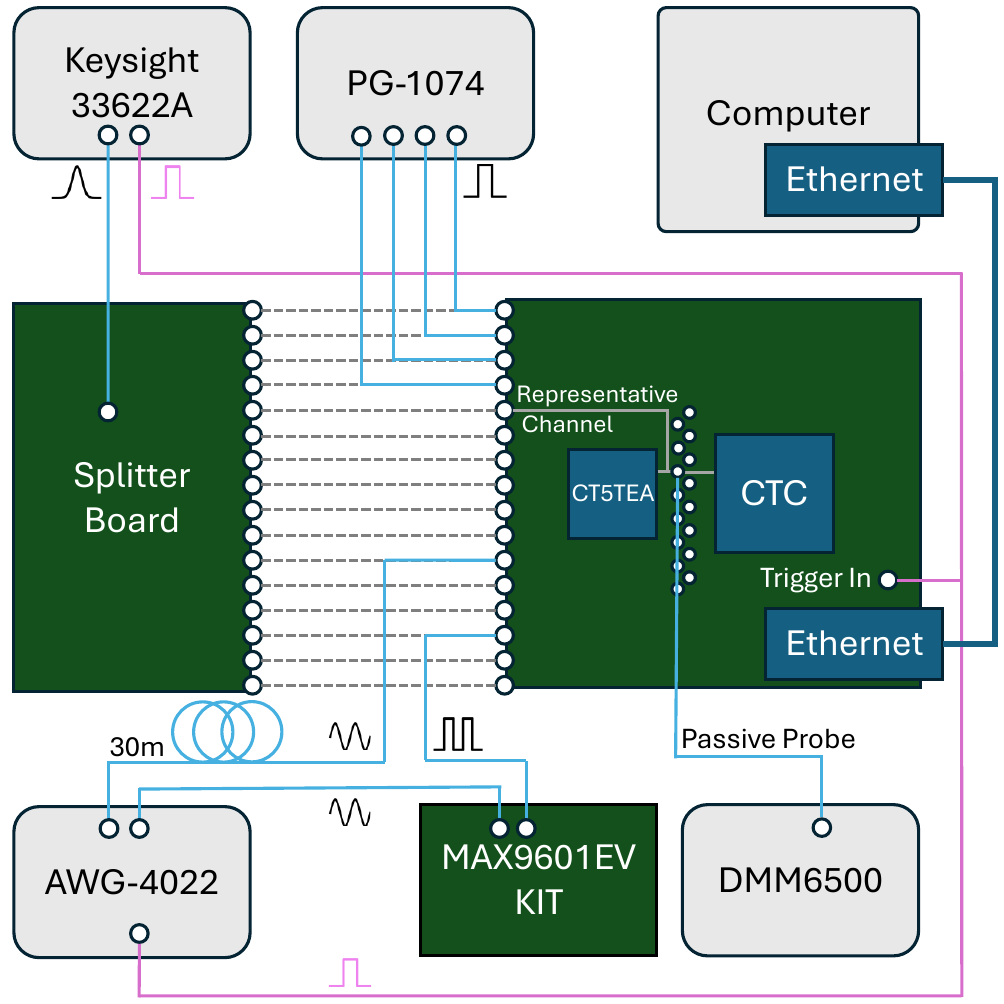}
\caption{\label{fig:SETUP} Schematic representation of all measurement setup configurations. Each function generation can be individually plugged into the inputs of the evaluation boards or connected to the splitter board to inject the signal in parallel to all 16 channels. However, only the sketched setups were used.}
\end{figure}

Figure \ref{fig:SETUP} shows a schematic representation of all measurement setup configurations. Each signal can be individually plugged into one of the inputs of the evaluation board or distributed by a custom splitter board based on operational amplifiers (Texas Instruments OPA692). As signal sources, several function generators were used: a Keysight 33622A function generator for Gaussian-shaped pulses and square trigger pulses, an Active Technologies PG-1074 for square pulses with a rise time of 70\,ps, an Active Technologies AWG-4022 for sine wave pulses, and a MAX9601EVKIT for translating such sine wave pulses to square waves. A digital multimeter (Keithley DMM6500) probes the DC offset at the ASIC inputs after the AC coupling. A detailed usage of each instrument is given in the specific section in which it is used.

The TARGET libraries \citep{ctagit}, a C++-based software, is used for communication with the FPGA and waveform data processing. It contains three packages for data readout and event building, as well as calibration of waveforms. \mbox{\textit{TargetDriver}} provides functions to configure the FPGA and TARGET ASICs by simple register operations (read/write) and higher-level functions, for example, initializing an evaluation board with default settings. Routines for a camera state machine, as well as the event builder, are also provided. \mbox{\textit{TargetIO}} handles the reading and writing of data. \mbox{\textit{TargetCalib}} contains the calibration routines for TARGET data. Basic functions are also available in Python wrappers using SWIG, enabling the use of simple Python scripts.

\section{Performance}
\label{sec:perf}
In the following section, the focus is set on the performance measurements of the ASICs.

\subsection{Power Consumption}

The power consumption can be monitored by measuring the voltage drop over an external 0.1\,$\Omega$ series resistor after the LDO linear regulator. Initialized and operating at a voltage offset of 750\,mV (see Section \ref{sec:DCTF}), the operational pedestal voltage, the pair of ASICs consume on average (942.7 $\pm$ 0.1)\,mW, resulting in a moderate power consumption of (58.9 $\pm$ 0.1)\,mW per channel. This value strongly depends on the selected operational pedestal voltage. The minimum is at the lowest settable pedestal voltage. As a common reference, the total consumption is \mbox{(641.7 $\pm$ 0.1)\,mW} at 150\,mV. The highest total consumption is at an operational pedestal voltage of 1360\,mV and is  \mbox{(1051.6 $\pm$ 0.1)\,mW}. A correlation of the power consumption with the trigger rate or the amplitude of input pulses is not observed.

\subsection{\label{sec:internaldac}Internal DAC performance}

\begin{figure}[t]
\centering
\includegraphics[width=0.48\textwidth]{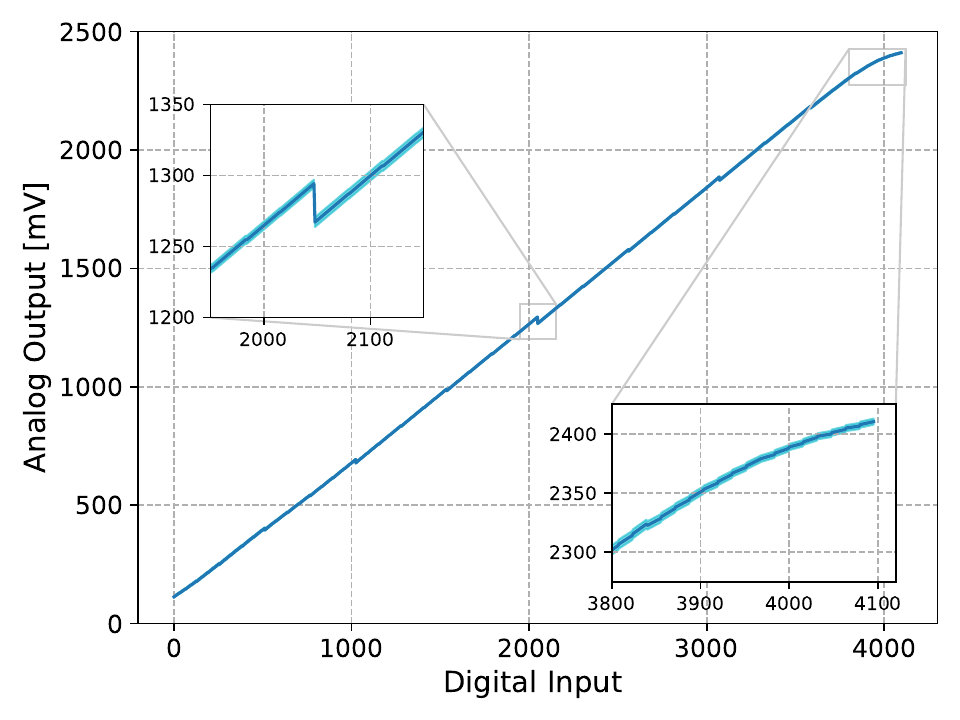}
\caption{\label{fig:VPED_example} The mean transfer function of the CT5TEA DACs of SN0001 and SN0002 controlling the pedestal voltage of the signal path with the standard deviation over all channels as semi-transparent area.}
\end{figure}

To operate and set up the ASICs, different bias voltages are needed across the whole functionality of the ASICs. These DC voltages are created by 12-bit R-2R resistor ladder networks. The resistors are tuned so that no voltage gaps occur in the DAC output voltage as a function of DAC input value (see Figure \ref{fig:VPED_example}). Most prominent of them are the CT5TEA ones that set up the necessary pedestal voltage per channel on the signal path, called \texttt{Vped}. These DAC output voltages were probed at the test point in front of CTC and measured with the Keithley DMM6500 multimeter with a mean standard error of $<$ 0.03\,mV over 512 different DAC settings. A measurement is taken before and after four bits flip simultaneously to intercept significant voltage drops in the transfer function due to inaccuracies of the R-2R resistor ladder network. The range and linearity of the DAC output voltage can be tuned by the \texttt{VpedBias} parameter regulating the bias current of the output amplifier of the DAC. A higher bias leads to an earlier saturation of the transfer function, reducing the dynamic range with no observable effect on the standard deviation. With current settings, an output voltage range of $> 2.2$\,V is achieved. The DAC performance is representative of all DACs in the TARGET ASICs.

\begin{figure}[t]
\centering
\includegraphics[width=0.48\textwidth]{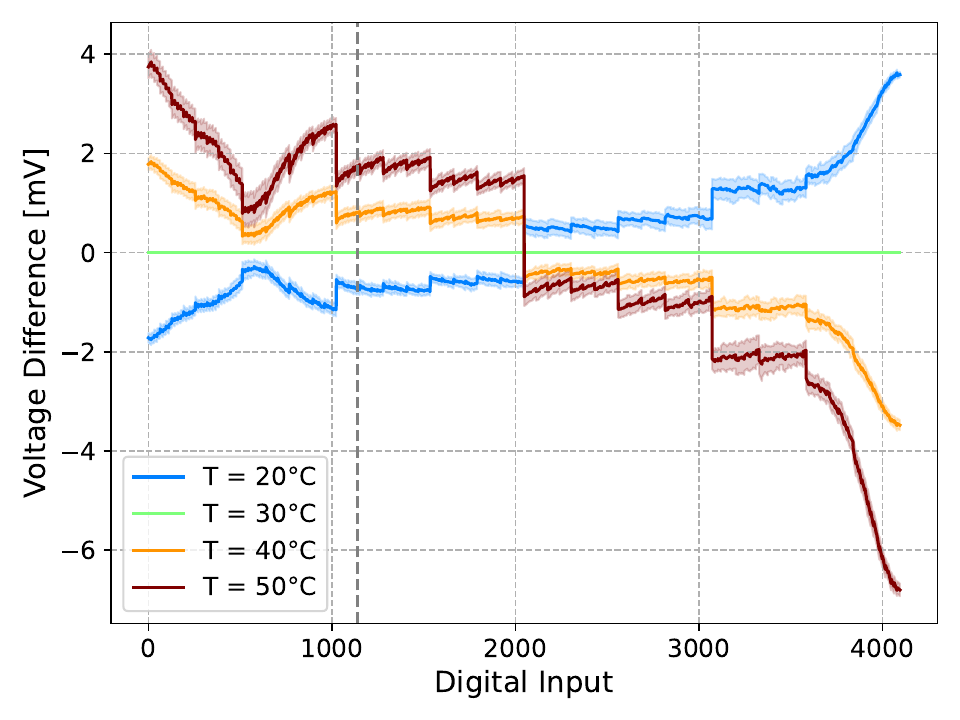}
\caption{\label{fig:VPED_T_dep} Temperature-induced voltage difference of the CT5TEA DAC transfer function per DAC setting of SN0002. Standard deviation between the 16 channels marked as semi-transparent area and the operational pedestal voltage marked as gray dashed line.}
\end{figure}

For SN0002, an analysis of the temperature dependence of the DACs was performed, as temperature in the SST camera will not be constant. The output voltage difference due to a change in temperature for a given DAC setting is presented in Figure \ref{fig:VPED_T_dep}. The steps arise from bit flips in the R-2R resistor ladder network, introducing different numbers of resistors with different temperature dependence to the circuit. Around the operational pedestal voltage of 750\,mV (gray dashed line), the temperature-induced offset scales with about 0.08\,$\frac{\mathrm{mV}}{\mathrm{K}}$ assuming a linear response. For the saturation regime, the temperature dependence increases; however, as a fraction of the input signal (which reaches 2.4 V at a DAC value of 4096), the effect is negligible. The same is true for DAC values below 800, as voltages below 600\,mV are not suitable for the CMOS SCA. For the use of an internal DC calibration (see Section \ref{sec:DCTF}), the DACs are deemed sufficiently temperature stable.

\subsection{Trigger Path (CT5TEA)}
In this section, the trigger performance of the CT5TEA is presented.

\subsubsection{Trigger output signal}

\begin{figure}[t]
\centering
\includegraphics[width=0.48\textwidth]{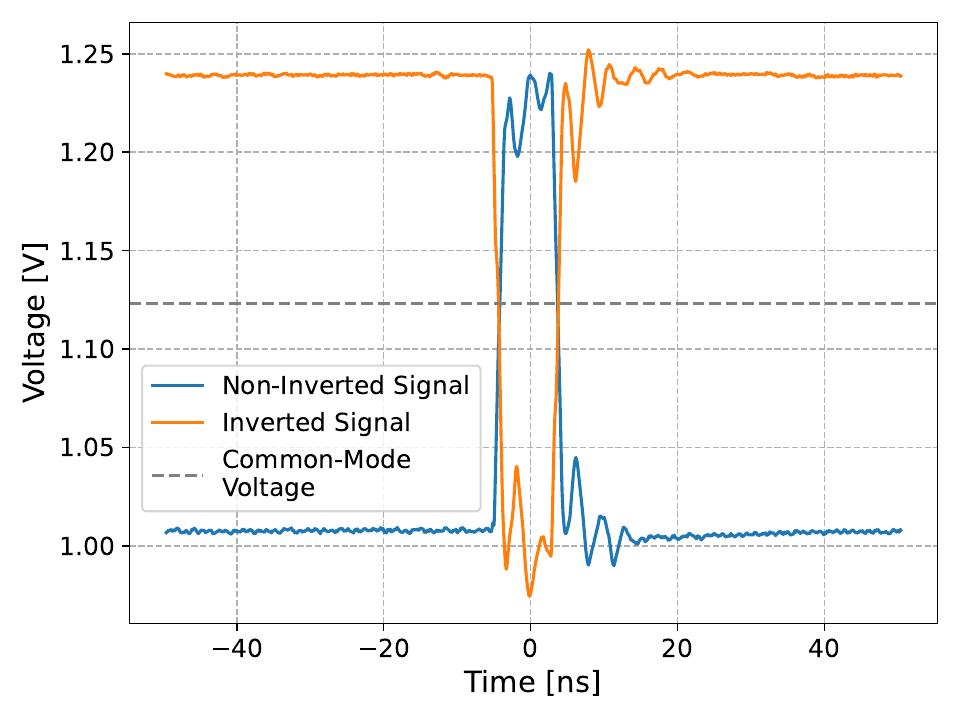}
\caption{\label{fig:triggerpulse} Average CT5TEA trigger pulse for a \texttt{WBias} setting of 1200 trigger pulse, probed with a Tektronix TDP7704 probe, soldered to the 100\,$\Omega$ termination resistor close to the FPGA. The extensive ringing is due to the limited bandwidth of the CT5TEA internal LVDS drivers.}
\end{figure}

The trigger of CT5TEA is based on the analog sum of four channels, forming four trigger groups per ASIC. A one-shot trigger pulse is generated if the sum exceeds a certain threshold, which can be adjusted by the parameters \texttt{Thresh} and \texttt{PMTRef4}. While \texttt{Thresh} sets the threshold applied to the comparator, \texttt{PMTRef4} changes the baseline of the summed signal. Therefore, setting the same effective trigger threshold with different pairs of \texttt{Thresh} and \texttt{PMTRef4} is possible. The width of the resulting trigger pulse is set via the \texttt{WBias} parameter. Figure \ref{fig:triggerpulse} shows the average waveform of 100 trigger pulses for a \texttt{WBias} of 1200, resulting in a width of roughly 8\,ns. These measurements use the evaluation board's ability to output the trigger pulses as single-ended signals via a coaxial connector. The FPGA does the LVDS-to-single-ended conversion. For the CT5TEA ASIC on SN0001, the spread in output pulse width is (19.2 $\pm$ 2.2)\,ps. The jitter between groups triggering on the same pulse is (53.3 $\pm$ 2.6)\,ps. The common mode voltage is (1126.6 $\pm$ 0.6)\,mV and the differential voltage swing is (474.4 $\pm$ 2.3)\,mV for the shown group 0. The trigger output signal complies with the Xilinx LVDS receivers \citep{Artix7}. All these measurements have been performed using the Tektronix MSO064b oscilloscope measurement functions.

\begin{figure}[h]
\centering
\includegraphics[width=0.48\textwidth]{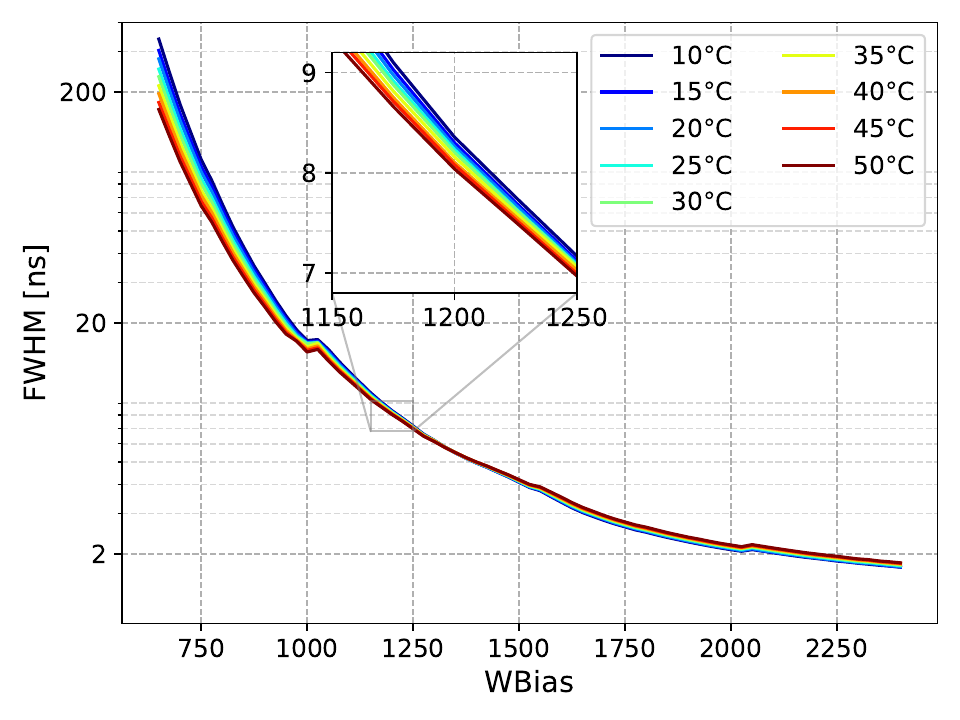}
\caption{\label{fig:trigger_width} FWHM of trigger signal vs \texttt{WBias} for different temperatures of module SN0001. The evaluation board FPGA was used to convert the LVDS signal into a single-ended signal on which the width measurement was performed. A difference of about 50\,ps is observed between direct LVDS and single-ended measurements for 2\,ns pulses, and up to 250\,ps difference for 15\,ns.}
\end{figure}

Figure \ref{fig:trigger_width} shows the width of the trigger signal as a function of the \texttt{WBias} setting, along with the variation in width with changing temperature. The jumps in the plot are again due to the design of the DACs as described in Section \ref{sec:internaldac}. The width ranges from $>$\,100\,ns down to $<$\,2\,ns, covering the 8\,ns specification for SST and 2\,ns specification for SCT \citep{Jama}. For a \texttt{WBias} of 1200, the pulse length scales linearly with \mbox{$\approx$ -6.7\,$\frac{\mathrm{ps}}{\mathrm{K}}$} for six of the eight tested trigger groups and \mbox{$\approx$ -17.7\,$\frac{\mathrm{ps}}{\mathrm{K}}$} for the other two. Higher statistics are needed to investigate the bimodal distribution. In the best case, a temperature change of 10\,$^\circ$C would cause a decrease of pulse length of $\leq$ 0.9\,\% for the 8\,ns pulse width and $\leq$ 2.3\,\% in the worst case.

\subsubsection{Effective trigger threshold and noise}
\label{sec:trigger_thresh_noise}

For different ASICs, the same threshold parameters result in different effective\footnote{The word "effective" is used to distinguish a threshold in mV from the \texttt{Thresh} parameter} thresholds. Therefore, scans of the \texttt{Thresh} and \texttt{PMTref4} parameter space are performed to generate a look-up table for operation. This procedure determines the range of available effective threshold values and the corresponding noise. For this purpose, pulses with variable amplitudes generated by the Keysight 33622A function generator are injected into the evaluation board using the custom splitter board described in Section \ref{sec:evaluationboard}. The outputs of the splitter board channels are calibrated to have a known amplitude as input to the evaluation board. For the measurements shown in this section, a Gaussian-shaped pulse with 10\,ns FWHM is used, similar to the shaped SiPM pulses in the SST camera. %In section \ref{sec:trigeffvswidth}, the influence of the pulse width on the effective trigger threshold is examined.

\begin{figure}[t]
\centering
\includegraphics[width=0.48\textwidth]{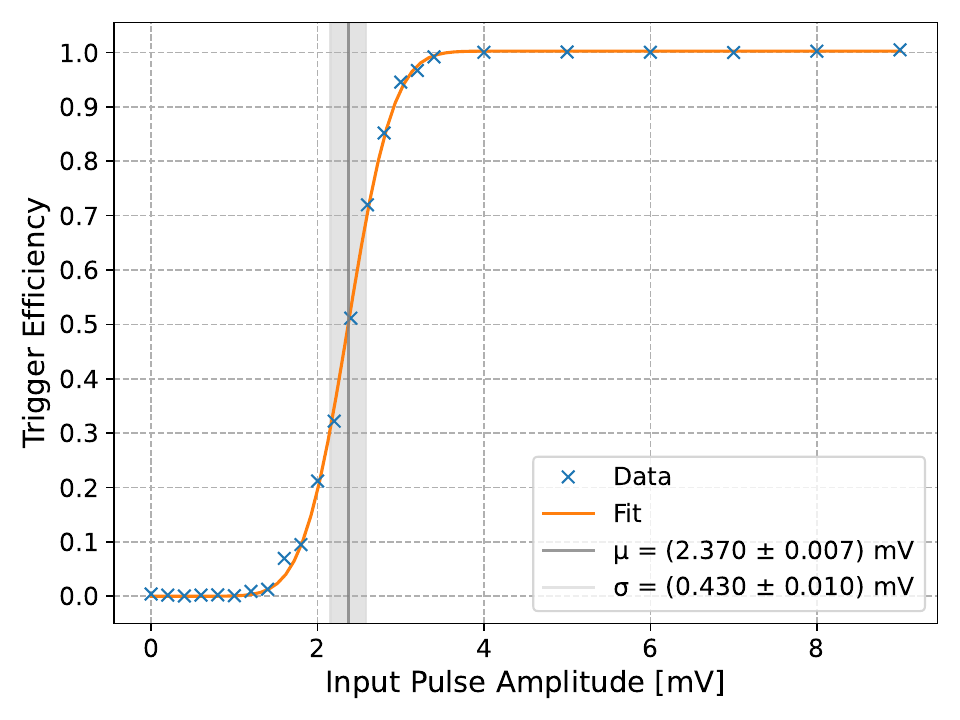}
\caption{\label{fig:trigger_scurve} Sigmoidal curve for a \texttt{Thresh} 2800 and \texttt{PMTref4} 2020 on SN0002. The input pulse amplitude is the sum of all four input channels of the trigger group.}
\end{figure}

The signal was injected into all four channels of each trigger group, while the amplitude in the following figures is the sum of all four channels. The gain of the individual channels can be fine-tuned using the \texttt{TriggerGain\_ChN} settings, where N represents the channel from 0 to 15. The trigger pulses are then counted via counters implemented in the FPGA. For each \texttt{Thresh}/\texttt{PMTref4} pair of the scan, the amplitude of the injected pulses is varied and the trigger pulse rate is measured as a function of the input amplitude. If the effective threshold is below the signal baseline, there will be no trigger pulses independent of the amplitude. If it is above the baseline, the rate will increase as soon as the pulse height matches the effective threshold and approaches 100\,\% of the input pulse rate as the pulse height increases beyond the effective threshold. Figure \ref{fig:trigger_scurve} shows such a sigmoidal curve for a specific parameter pair. The values for the effective threshold $\mathrm{\mu}$ and the trigger noise $\sigma$ are determined by fitting an error function of the form in Equation \ref{eq:error} to the data.

%\begin{align}
\begin{equation}
\label{eq:error}
    S(x,\mu,\sigma) = \frac{1}{2} \left(1 + \mathrm{erf}\left( \frac{x- \mu}{\sqrt{2}\sigma} \right) \right) 
\end{equation}
%\end{align}

\begin{figure}[h]
\centering
\includegraphics[width=0.48\textwidth]{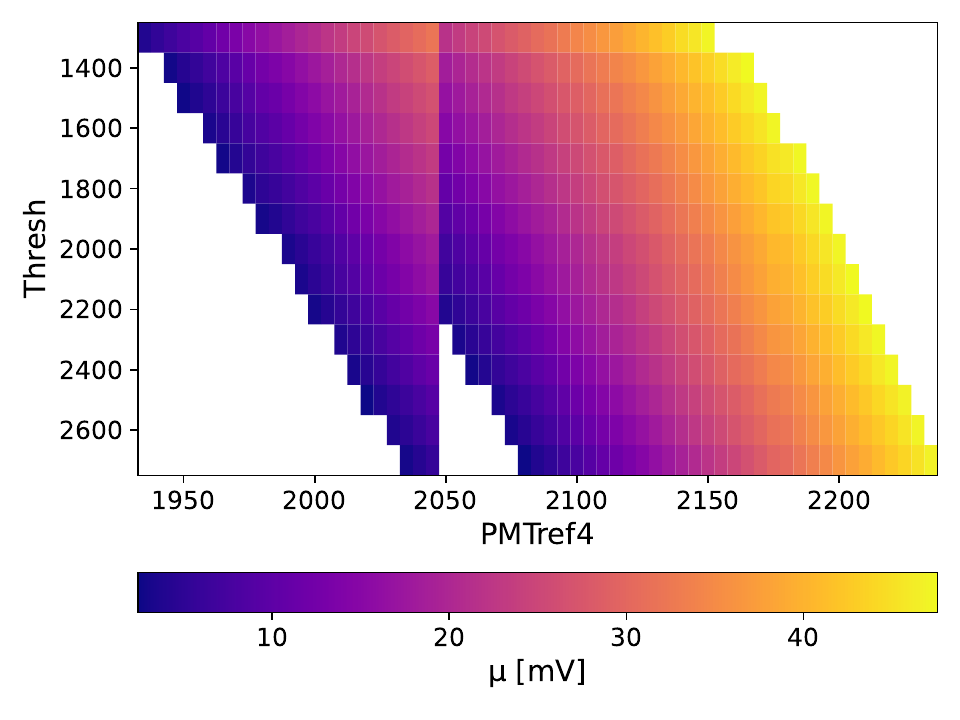}
\linebreak
\includegraphics[width=0.48\textwidth]{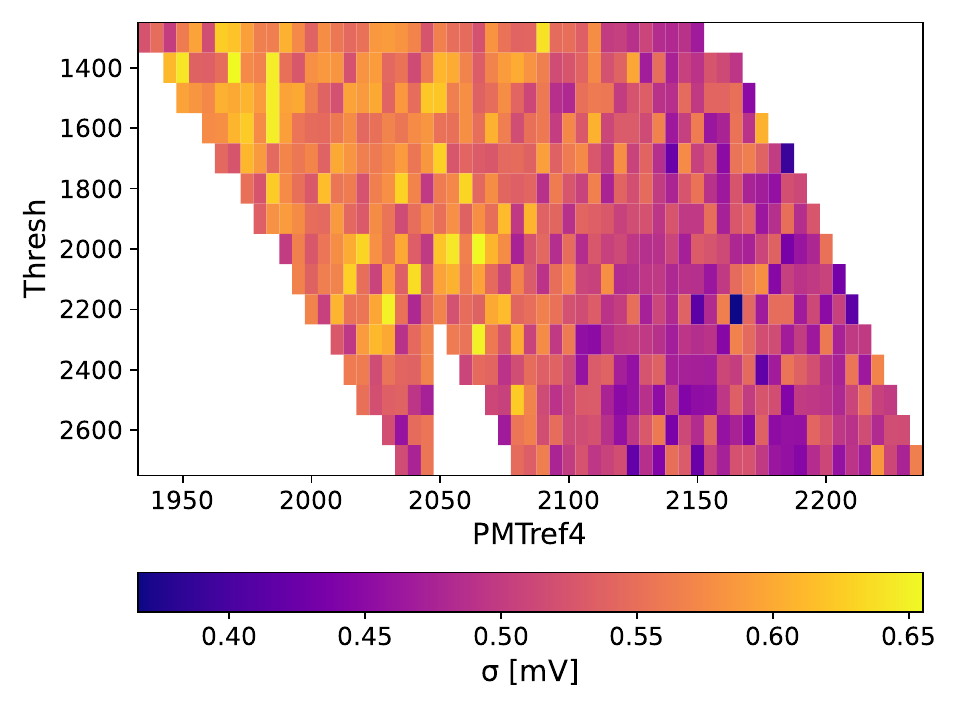}
\caption{\label{fig:trigger_map} Top: Effective threshold for SN0002 and different pairs of \texttt{PMTref4} and \texttt{Thresh}, the jump at 2048 is caused by the architecture of \texttt{PMTref4} DAC. See Section \ref{sec:internaldac} for details. Bottom: Corresponding noise.}
\end{figure}

In Figure \ref{fig:trigger_map}, a 2D map of the effective threshold and noise as a function of DAC settings \texttt{Thresh} and \texttt{PMTref4} is shown. Coarser scans were performed to verify that effective thresholds of $>$ 250\,mV can be achieved for all groups. As the \texttt{Thresh} DAC sets the second input of the trigger comparator after amplification, while the \texttt{PMTref4} DAC directly adjusts the offset, the effect on the effective threshold from changing \texttt{PMTref4} is larger.

\begin{figure}[t]
\centering
\includegraphics[width=0.48\textwidth]{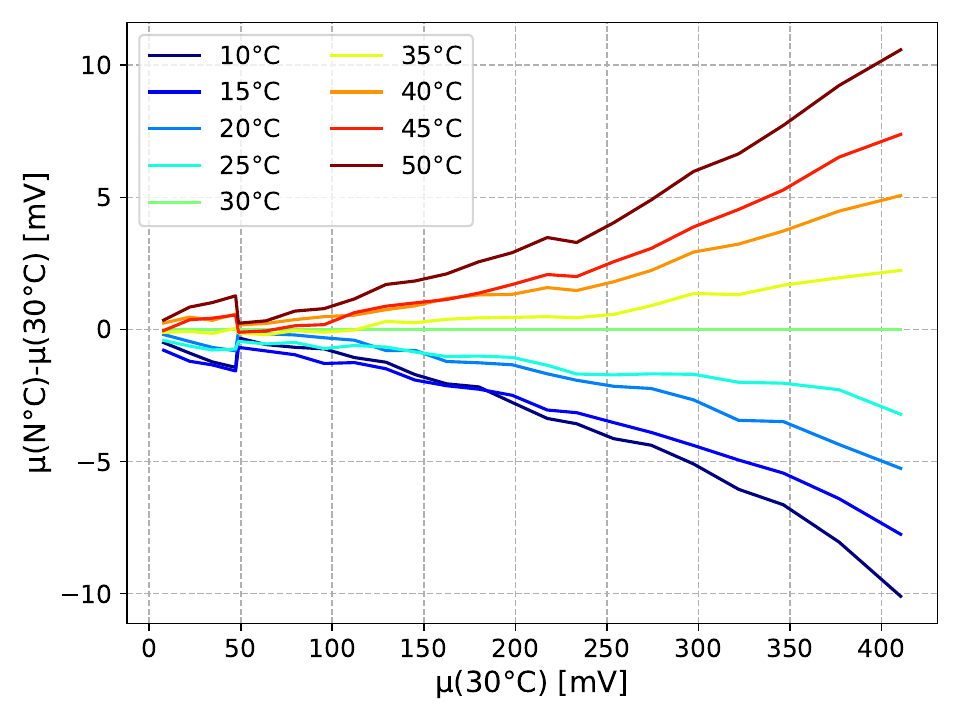}
\caption{\label{fig:trigger_tempdep} Temperature dependence of the effective threshold for group 1 in SN0001. The jump at 50\,mV originates from a change \mbox{$<$ 2048} to \mbox{$>$ 2048}  in \texttt{PMTref4}.}
\end{figure}

It is important to note that all these characterization measurements were performed with sampling running, as this resulted in increased noise in the previous ASIC generations and was the main reason for separating CT5TEA/CTC into two ASICs. For the eight tested trigger groups, at temperatures between 10\,$^{\circ}$C and 50\,$^{\circ}$C, the average minimum effective threshold is (2.80 $\pm$ 0.15)\,mV varying between 2.4\,mV and 3.25\,mV. The corresponding noise is (0.51 $\pm$ 0.04)\,mV. If a larger fraction of noise triggers were acceptable, the effective threshold could decrease further. For a threshold between 10\,mV and 40\,mV, the average noise is (0.52 $\pm$ 0.16)\,mV.
%Do you mean "If a larger fraction of noise triggers were acceptable, the effective threshold could decrease further"?
To complete the picture, the temperature dependence of the effective threshold is examined. Figure \ref{fig:trigger_tempdep} shows the result for the first trigger group of SN0001. A 1D look-up table was generated at 30\,${^\circ}$C. The same parameter pairs were then tested at different temperatures, and the 30\,$^{\circ}$C reference subtracted from the results. With higher temperatures, the effective threshold increases. The jumps observed in temperature dependency correspond to the jumps in the transfer function of the DACs, as seen in Section \ref{sec:internaldac}. For small effective thresholds, the temperature dependence is $\leq$ 0.1\,$\frac{\mathrm{mV}}{\mathrm{K}}$ and therefore negligible, while for larger amplitudes, the temperature-induced offset can grow up to 0.5\,$\frac{\mathrm{mV}}{\mathrm{K}}$.% For the more unlikely case of a 100\,p.e. effective threshold, assuming 3.37\,$\frac{\mathrm{p.e.}}{\mathrm{mV}}$, a 10 degree shift would therefore shift the effective threshold up about 1.5\,p.e..

\subsubsection{\label{sec:trigeffvswidth}Trigger efficiency as a function of pulse width}

\begin{figure}[t]
\centering
\includegraphics[width=0.48\textwidth]{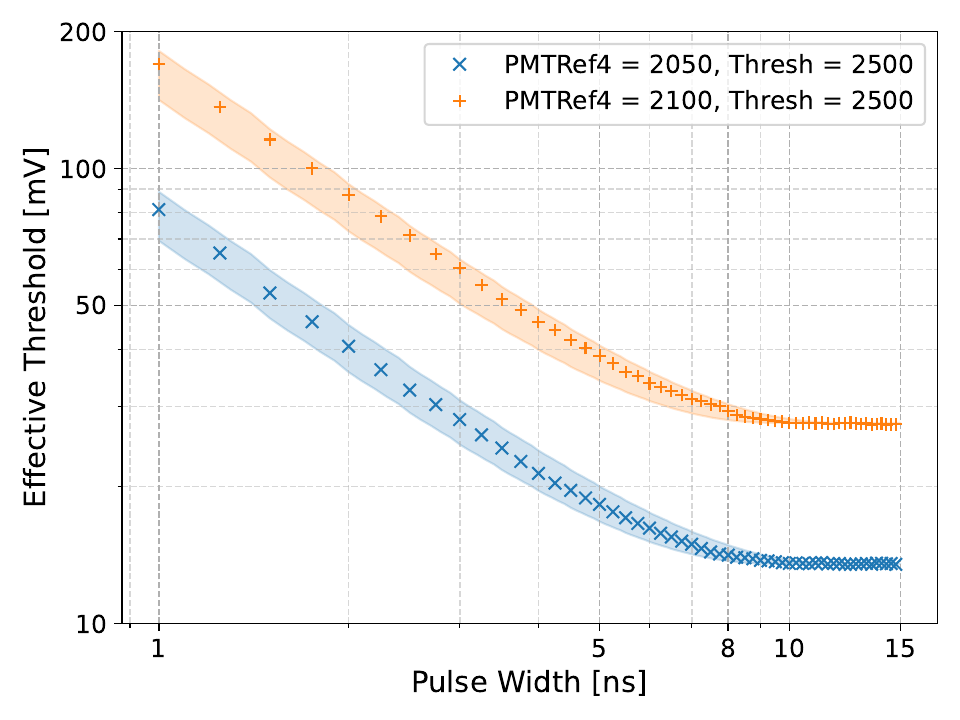}
\caption{\label{fig:twidth_vs_thresh} The effective threshold for a rectangular pulse with different FWHM for two parameter pairs. The colored bands represent the amplitude required for a pulse of specific width passing through a fourth-order Butterworth filter with a cutoff frequency between 70 and 80\,MHz to achieve the same amplitude as a 15\,ns FWHM pulse.}
\end{figure}

This section investigates the influence of the input pulse width on the effective threshold. For this purpose, the four outputs of the Active Technologies PG-1074 pulse generator are connected to the four input channels of a trigger group. The pulse generator produced rectangular pulses with variable amplitude and width, with a rise-time of 70\,ps. For different pulse widths in the range of 1\,ns to 15\,ns FWHM, the efficiency scans of the previous sections are done for two different \texttt{Thresh}/\texttt{PMTref4} pairs. The result is presented in Figure \ref{fig:twidth_vs_thresh}. Above 10\,ns FWHM, the effective threshold does not change. Below, it exponentially increases with decreasing width. This is explained by the limited bandwidth of the CT5TEA internal amplifiers. The bands in the plot represent the amplitudes required for an input pulse to attain the equivalent amplitude of a 15\,ns FWHM pulse, both processed through a fourth-order Butterworth filter \citep{butterworth1930} with a cut-off frequency ranging from 70 to 90\,MHz. This frequency matches the 10\,ns Gaussian-shaped pulse and the bandwidth of the shaping amplifiers in the SST camera well.

Based on the results presented in this section, it is concluded that the CT5TEA is suitable for use in Cherenkov cameras and for similar applications.

\subsection{Data Path (CTC)}
In this section, the performance of the CTC ASIC is presented. 

\subsubsection{Sampling and time base}
\label{sec:time_base}

Based on the earlier work on the TARGET 5 ASIC \citep{TARGET_5_paper}, the sampling frequency is set to the default value of 1\,GSa/s and is kept unchanged for all tests. While the beginning of every sampling cycle is synchronized with the FPGA clock, the length of the individual sampling cell has to be fine-tuned, such that the transition from the last cell of the second block to the first cell of the first block is in sync with the FPGA clock. The total length of the sampling array can be tuned with two settings. Via \texttt{SSToutFB\_Delay} one can tune the length of the sampling array with 1\,ns (at 1\,GSa/s) granularity over a large range. By changing the \texttt{VtrimT} parameter, the length is controllable on the sub-ns level to fine-tune the total length to 64\,ns.

A \texttt{VtrimT} scan is performed for each board at a constant \texttt{SSToutFB\_Delay}. Here, a 50\,MHz sine wave is applied to a channel for different \texttt{VtrimT} values. The digitized waveform of 128\,ns length is then separated into two parts, split at the transition of sampling cell 63 to 0. Both parts are then individually fitted with a sine wave. The resulting phases of both fits are subtracted from each other. With a correctly set \texttt{VtrimT}, a phase difference below 20\,ps is achieved, which is more than sufficient for operation.

To study the width of each sampling cell, square waves are used. The normalized distribution of edge transitions in the cells divided by the sampling frequency yields the width of each cell. This method was used instead of measuring zero-crossings of a sine wave \citep{zero_crossing}, as sharp edges of square waves yield more precise results that are less sensitive to amplitude miss-calibration. To generate a square-wave signal with a rise time (10 - 90\%) of 200\,ps and an amplitude of about 200\,mV, a sine wave generated by the AWG-4022 function generator is fed into the MAX9601EVKIT board, an ultra-fast comparator. The edge of the comparator output pulse is then assigned to the sampling cell if the previous sample is under a certain threshold and the sample itself and the following one are above it. This threshold is arbitrary for sufficiently fast rise times and was set to 100\,mV. An external trigger was used to cover all sampling cells equally, and 100,000 waveforms were recorded.

\begin{figure}[t]
\centering 
\includegraphics[width=0.48\textwidth]{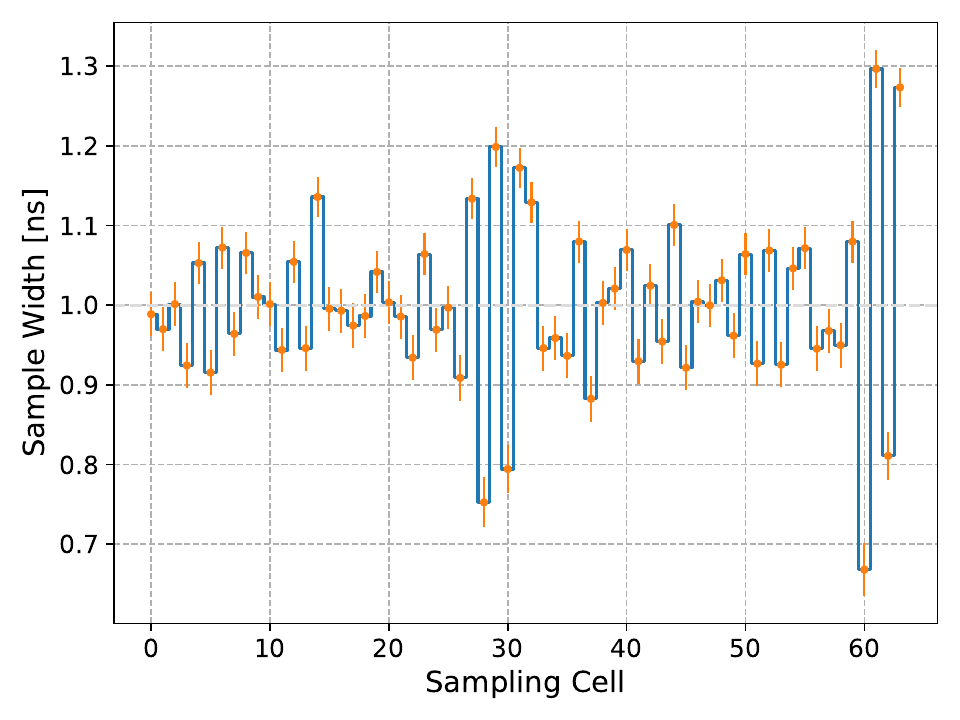}
\caption{\label{fig:TimeBase} Individual sampling cell width of the sampling array of \mbox{channel 0} of SN0001 with the corresponding standard error.}
\end{figure}

The time base for channel 0 of SN0001 after \texttt{VtrimT} tuning is shown in Figure \ref{fig:TimeBase}. The spread around 1\,ns is about 60\,ps in standard deviation, excluding the 28th to 32nd cells of each block which further deviate up to 300\ ps. It will later be seen that these cells show a different behavior in a range of other aspects. It is assumed that this behavior is due to the physical layout of the ASIC substrate, but a clear single indicator of why these cells are different was not found.

No shift of the time base between the channels is observed, which is a fair assumption as the time base signal propagates from channel 0 to channel 15. For the rest of this work, the time base is treated as 1\,ns bin. Pulse shape distortion caused by the uneven sampling time base is handled by different calibration techniques and ultimately increases the effective AC noise (see Section \ref{sec:AC_noise}).

\subsubsection{DC transfer function and noise}
\label{sec:DCTF}
Individual storage cells have slightly different capacitor sizes that are read by non-linear Wilkinson ADCs. Thus, each storage cell needs to be calibrated against different input voltages to characterize the relation between voltages and ADC values (transfer function).
This is done internally on the board with the DACs of CT5TEA creating the pedestal voltages as reference. Therefore, it is possible to re-calibrate the DC transfer function of the ASICs in situ. For on-the-bench calibration, an external trigger is used to cover all storage cells equally. For this purpose, the Keysight 33622A function generator is used, creating a 3.3\,V square pulse of 25\,ns width at 600\,Hz, connected to the board's trigger input. The clocks of the function generator and the evaluation board are not synchronized to ensure an equal event distribution over all storage cells. For one DC transfer function, the full voltage range of the DACs of CT5TEA is used in steps of 50\,mV, resulting in a maximum voltage range of about 2.2\,V.

To access the maximum voltage range of CTC, the comparator of the Wilkinson ADCs had to be tuned. This was done on SN0001 and later adopted by SN0002. The slope and resulting resolution of the DC transfer function is determined by the slope of the Wilkinson ramp, which is generated by charging an external ramp capacitor with an adjustable constant current, controlled internally by \texttt{Isel}. The discharge (zero) level can be set via the \texttt{Vdischarge} parameter. The ramp and zero level is tuned to create the maximal ADC count to voltage ratio over the full voltage range with a small margin to account for ASIC-to-ASIC and temperature variations. As the slope of the ramp is the determining parameter, the slope of SN0001 is used as a reference to adjust the \texttt{Isel} of each ASIC by comparing the ramp monitor outputs, which delivers a buffered copy of the ramp. For SN0001 and SN0002, a mean effective bit range of 11.7 and 11.6 bits is achieved respectively.

For each DC transfer function, 20,000 waveforms per DC voltage are recorded. It has been found that the digitization phase\footnote{Position of the block in which the sample is located during the digitization process} of a storage cell influences the digitized value. Depending on whether the number of the digitization block is even or odd, the digitized value of the storage cell can change by up to $\pm$\,2\,mV from the mean over all digitization blocks. Therefore, for the standard operation of 128-sample waveforms, an individual transfer function is constructed for each storage cell and its five possible digitization phases. It should be noted that switching noise associated with the addressing of the storage array means that a given set of DC transfer functions are only valid for a set trigger delay, i.e. look-back time between receiving an trigger  and location in the storage array.

\begin{figure}[t]
\centering
\includegraphics[width=0.48\textwidth]{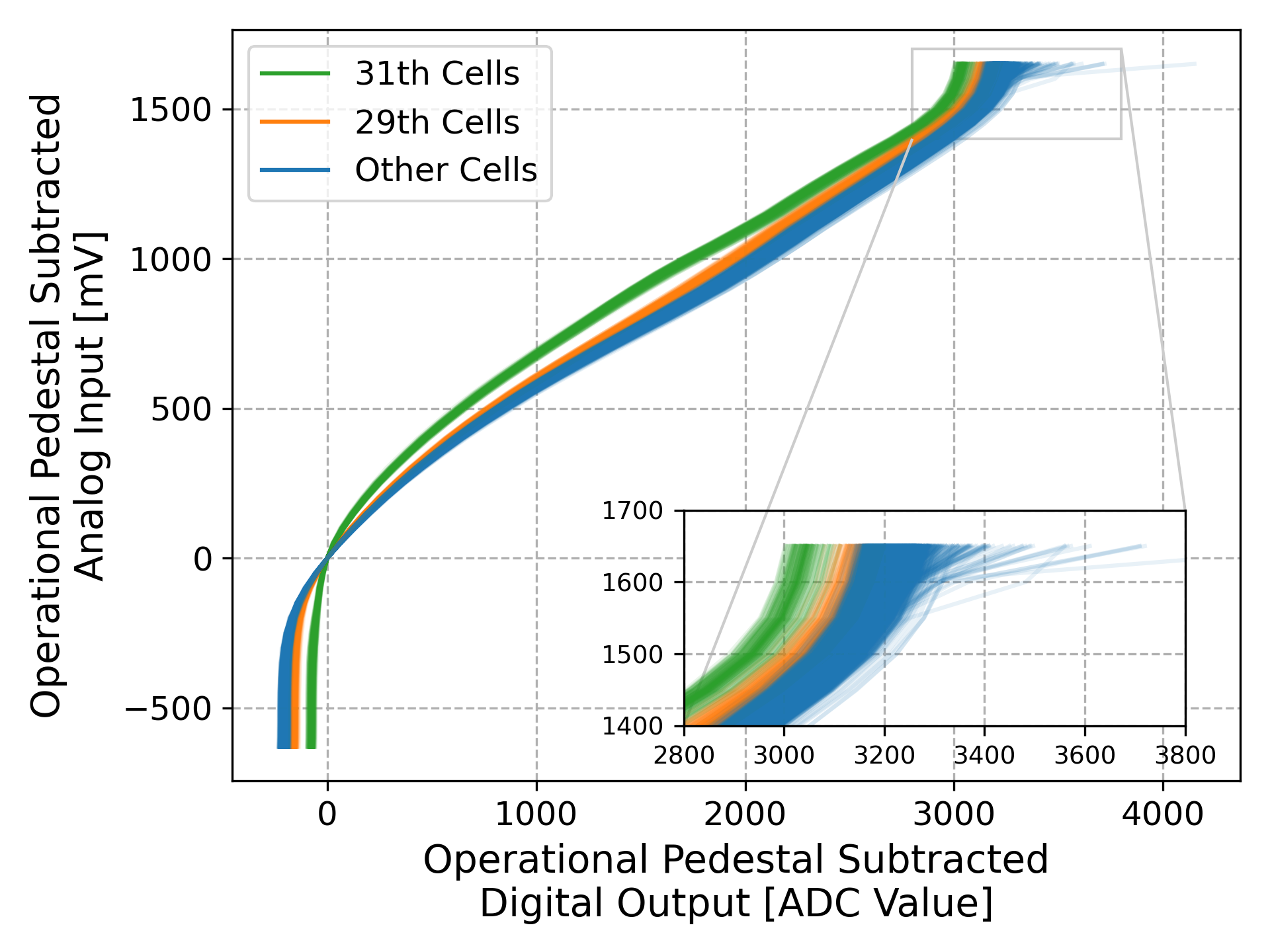}
\caption{\label{fig:DC_TF_example} All 4096 individual DC transfer functions of channel 0 for one digitization block phase of SN0001 with the operational pedestal already subtracted. The different behavior of each cell is color-coded, corresponding to its position in the storage block. A standard error or standard deviation of each individual transfer function is not given due to visibility reasons.}
\end{figure}

%The DC transfer functions are only valid for a certain trigger delay as there is different noise at certain time intervals caused by the switching noise of the storage array addressing.
Example DC transfer functions of SN0001 of channel 0 in the first digitization phase can be seen in Figure \ref{fig:DC_TF_example}. It is apparent that the operating pedestal voltage is already subtracted and that there are three distinct cell behaviors. This is a known feature since TARGET 5.

For large ADC counts, the slope/resolution of the transfer function increases/decreases again for most cells, with the exception of certain cells. These can be observed in the zoom panel of Figure \ref{fig:DC_TF_example}. Due to tolerances in the fabrication process of the ASICs, some storage cells saturate earlier. Consequently, the stored voltage rises slower than the applied analog input, leading to higher ADC count values for the same analog input.

\begin{figure}[t]
\centering
\includegraphics[width=0.48\textwidth]{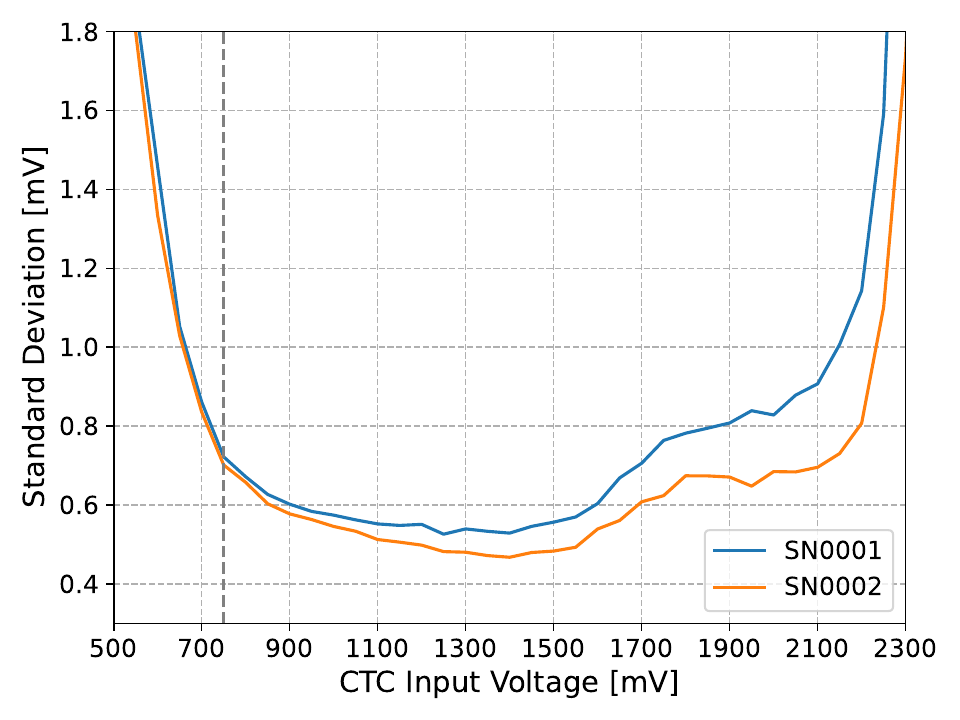}
\caption{\label{fig:DC_noise} DC noise level of each ASIC for the relevant voltages. The steep increase in noise to the edges results from the low mV to ADC ratio.}
\end{figure}

The DC noise (standard deviation of the waveform) corresponding to the above transfer function is shown in Figure \ref{fig:DC_noise}. Towards the increasing slopes of the DC transfer function, the DC noise increases drastically as the mV-to-ADC ratio drops significantly. The operational pedestal voltage of 750\,mV is therefore chosen as a compromise between the lowest possible baseline noise while retaining a high dynamical range of positive voltages.

The temperature within the SST camera is unlikely to remain uniform and constant over all operational conditions, and therefore a temperature analysis of the DC transfer functions has been performed. For each temperature, a DC transfer function with baseline subtraction is recorded. The baseline changes by up to 0.2\,$\frac{\mathrm{ADC}}{\mathrm{K}}$ and even 0.5\,$\frac{\mathrm{ADC}}{\mathrm{K}}$ for the 31st cell. But just like the DC transfer function, the baseline subtraction can be redone at any time, correcting the baseline shift independent from slope shifts.

\begin{figure}[t]
\centering
\includegraphics[width=0.48\textwidth]{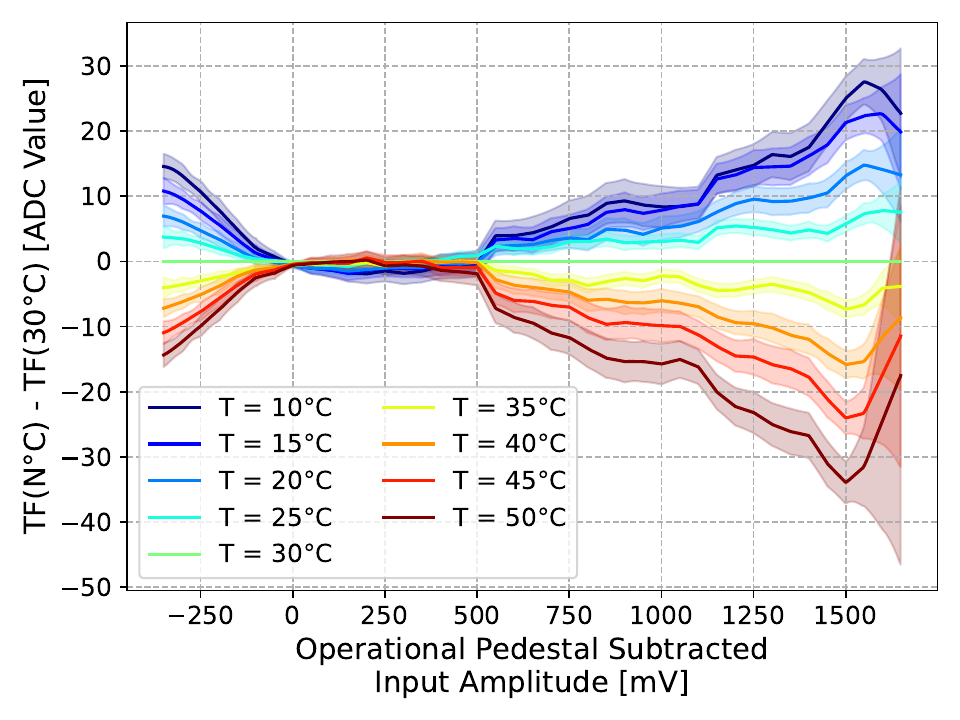}
\caption{\label{fig:DC_TF_T_dep}  Temperature dependence of a DC transfer function as voltage difference between the transfer functions at different temperatures of SN0001. Standard deviation over all channel and storage cells given as semi-transparent band.}
\end{figure}

An example of the temperature variations in the DC transfer function is presented in Figure \ref{fig:DC_TF_T_dep} for SN0001, where each transfer function over all channels, storage cells, and block phases at a certain temperature is subtracted from the corresponding transfer function at the reference temperature of 30\,$^\circ$C and then averaged, standard deviation as a semi-transparent band. From 0\,mV up until 500\,mV, the temperature dependence is nearly negligible with a change of about $<$\,0.1\,$\frac{\mathrm{ADC}}{\mathrm{K}}$. Above 500\,mV and below 0\,mV, the difference for one temperature grows approximately linearly with the pedestal voltage until the slope of the transfer function increases. With a temperature dependence of 1.75\,$\frac{\mathrm{ADC}}{\mathrm{K}}$, re-calibration of the DC transfer function is favorable for different temperatures.

\subsubsection{Bandwidth and cross-talk}

For the bandwidth and cross-talk measurements, one data set is used for each evaluation board. This set is created by applying a sine wave pulse of 400\,ns length, from 10\,MHz up to 600\,MHz at an amplitude of 500\,mV to each input channel individually with the Active Technologies AWG-4022 function generator over a 40\,m long coaxial cable. The short pulse and the long cable ensure that the amplitude of the sine wave signal is not distorted by reflections of mismatched input impedance of the evaluation board or ASIC. Additionally, the short pulse avoids memory effects in the storage cell array due to a possible charge-up of the individual capacities from a continuous sine wave. The operational pedestal voltage is set to 1.5\,V instead of 750\,mV to measure the negative amplitudes at a reasonable resolution. All data is calibrated with a DC transfer function with the 1.5\,V pedestal voltage offset. The amplitude is extracted through a sine wave fit. Although frequencies above 500\,MHz are not extractable for a 1GSa/s sampling, knowing the input frequency, the amplitude can still be extracted by a fit.

\begin{figure}[t]
\centering
\includegraphics[width=0.48\textwidth]{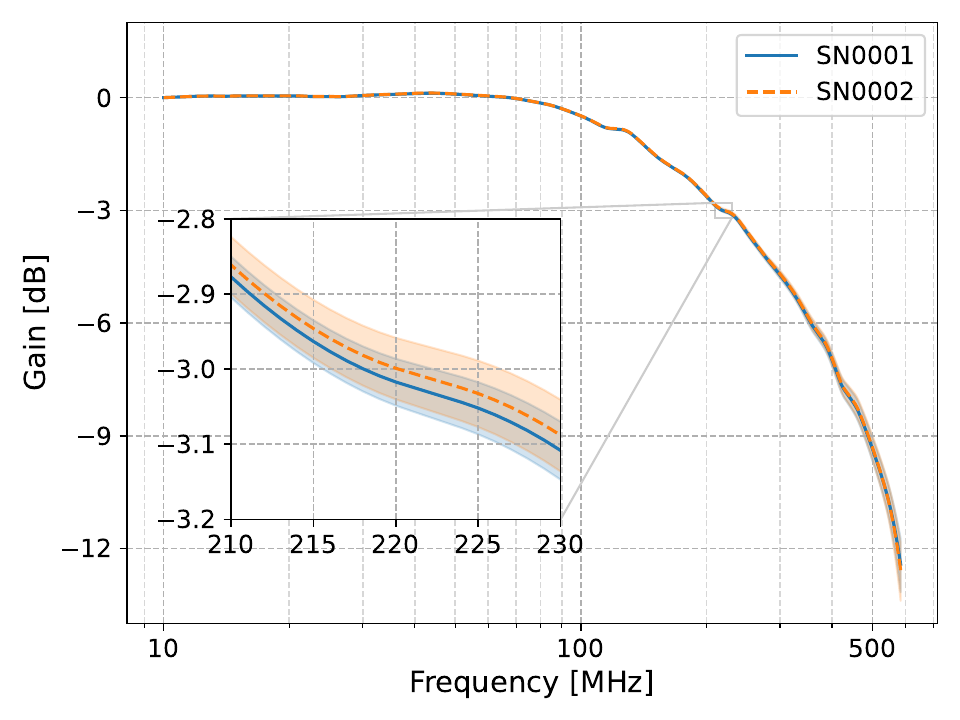}
\caption{\label{fig:bandwidth} The mean bandwidth of both ASICs with a 15\,MHz sliding average, - 3\,dB point highlighted. The standard deviation of the channel-to-channel spread as semi-transparent band.}
\end{figure}

The bandwidth of both boards can be seen in Figure \ref{fig:bandwidth} smoothed with a 15\,MHz sliding average to suppress ripples generated by the 64\,ns sampling cell array. The channel-to-channel standard deviation is given as a semi-transparent band. The - 3\,dB point is around 220\,MHz, with a gain flatness of 0.2\,dB up to 80\,MHz.

Taking the 5\,$\sigma$ width of a Gaussian-shaped pulse in the frequency spectrum, the ASIC is suited for pulses down to 3.6\,ns over the whole - 3\,dB bandwidth and digitizes pulses with widths up to 10\,ns undistorted. This is matched to the pulses in the SST camera, where the exponential decay of a SiPM is shaped into a pulse of 10\,ns width.

\begin{figure}[t]
\centering
\includegraphics[width=0.48\textwidth]{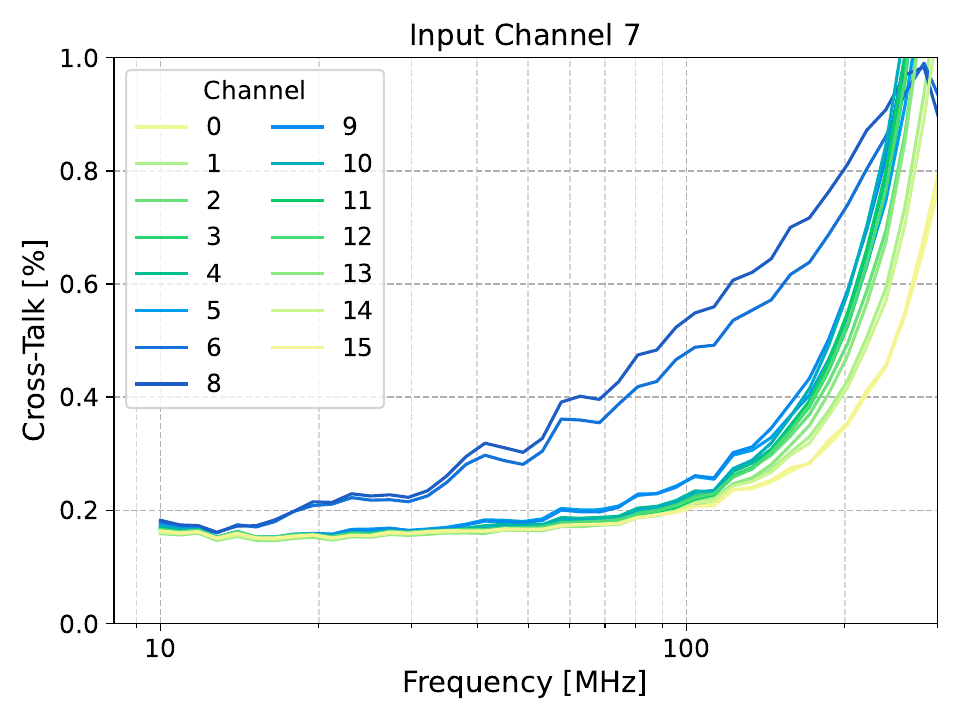}
\caption{\label{fig:Crosstalk_freq} Frequency dependent cross-talk with channel 7 as input channel. It should be noted that standard errors are included, but are not visible. The baseline noise of each channel creates an artificial cross-talk of about 0.2\,\%.}
\end{figure}

For the evaluation of the frequency-dependent cross-talk, the RMS of every measured channel divided by the RMS of the input channel is used. The results for channel 7 of SN0001 can be seen in Figure \ref{fig:Crosstalk_freq} with a mean standard error $<$0.01\,mV. The color gradient gives the distance to the input channel on the board. Only the next neighbor channels receive a measurable cross-talk, while the other channels simply weigh the baseline noise against the amplitude, resulting in an artificial cross-talk of 0.2\,\%. Therefore, the cross-talk grows strongly over 100\,MHz as the bandwidth of the ASIC drops, while the noise floor remains unaffected. Over the whole - 3\,dB bandwidth of the ASIC, the overall cross-talk does not exceed 1\,\% for both modules.

\begin{figure}[t]
\centering
\includegraphics[width=0.48\textwidth]{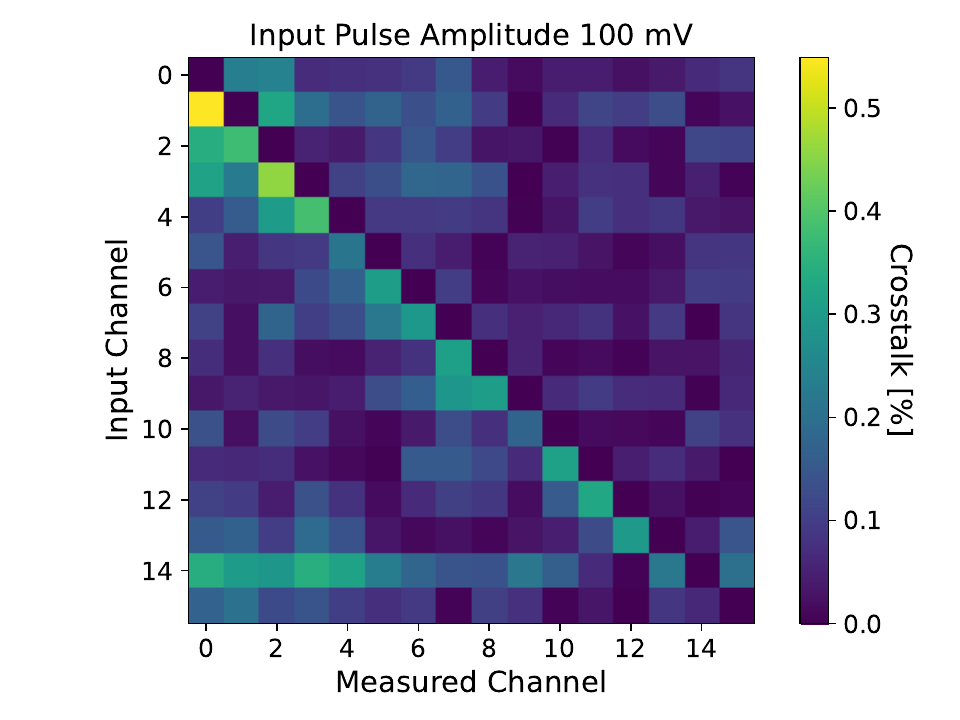}
\linebreak
\includegraphics[width=0.48\textwidth]{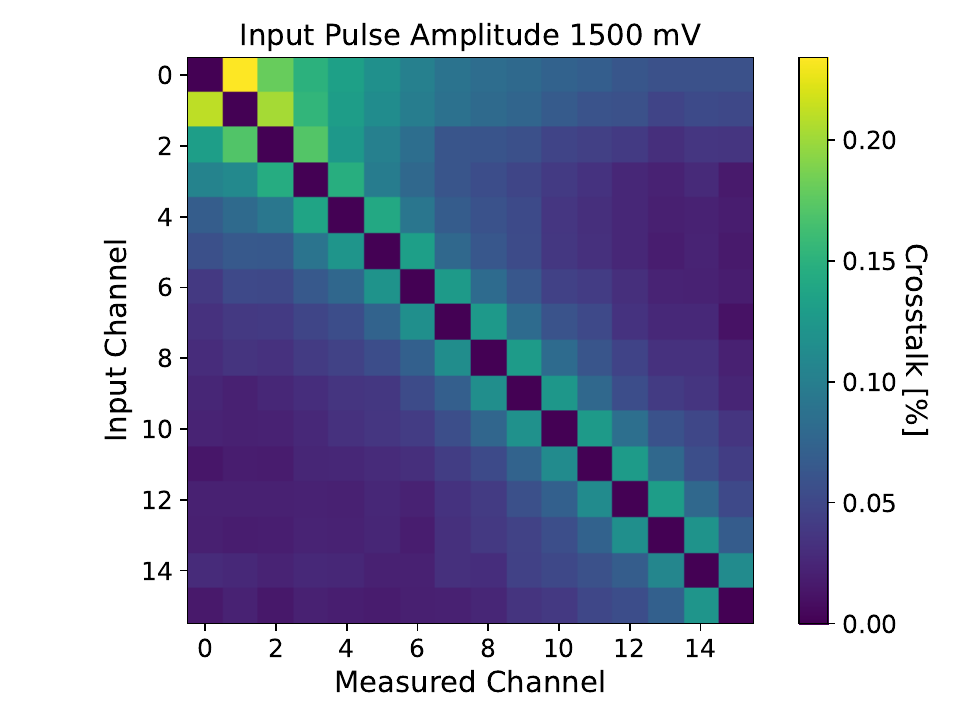}
\caption{\label{fig:Crosstalk_matrix} Cross-talk matrices for pulse amplitudes of 100\,mV (top) and 1.5\,V (bottom) of SN0001. The integral of the selected channel is divided by the integral of the input channel.}
\end{figure}

For a more realistic approach that mimics cross-talk as in the SST camera, Gaussian-shaped pulses of 10\,ns width are fed into the input channel. To quantify the level of cross-talk, the integrated pulse of the selected channel is divided against the integrated pulse of the input channel. The advantage of this method is that the integral minimizes the contribution of the noise floor, allowing a better signal-to-noise ratio and, therefore, less distorted cross-talk, as it is the favored method of charge extraction. In Figure \ref{fig:Crosstalk_matrix}, the corresponding cross-talk matrices for two different amplitudes can be seen. With the ground plane as close as possible to the signal track and a signal track distribution over two layers of the PCB, the cross-talk is successfully minimized to levels below $0.5\,\%$ for small amplitudes and 0.25\,\% for large amplitudes. The clustering of channel 0 to 3, and 15 and 16, suggests that most of the cross-talk is due to the traces of the evaluation board and not the ASIC itself, as these groups connect separately on top and bottom of the CT5TEA as opposed to the rest. This behavior is also visible in the frequency-dependent cross-talk.

\subsubsection{Effective AC Noise\label{sec:AC_noise}}

Most experiments using the TARGET ASICs receive pulsed or general AC data. Therefore, it is crucial to determine the effective AC noise of CTC. To measure the effective AC noise of the ASIC, the splitter board, introduced in Section \ref{sec:evaluationboard}, injects sine wave pulses of different frequencies with amplitudes between 1\,mV and 1000\,mV into each channel. For that purpose, the Keysight 33622A is used, also generating the trigger at 600\,Hz to cover all storage cells. 12,000 waveforms were recorded for each amplitude and each waveform is calibrated with a block-dependent DC transfer function. The data is fitted with a sine wave, and the standard deviation of the residuals yields a quantitative measurement of the effective AC noise. 

\begin{figure}[t]
\centering
\includegraphics[width=0.48\textwidth]{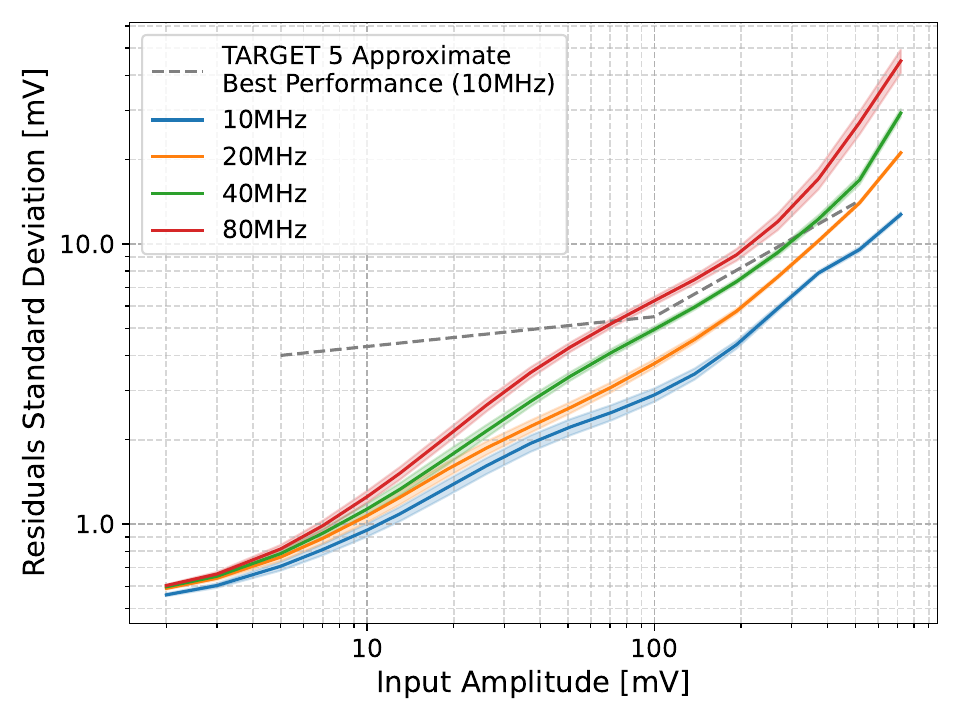}
\caption{\label{fig:AC_noise} AC noise of SN0001, with the standard deviation over each channel seen as a semi-transparent area. An approximated reference of the AC noise of  TARGET 5 is given as the gray dashed function.}
\end{figure}

The results for SN0001 are shown in Figure \ref{fig:AC_noise}, with the standard deviation between the different channels as a semi-transparent band. As a comparison, a linear approximation of the effective AC noise at 10\,MHz of TARGET 5 \citep{TARGET_5_paper} is also shown. The effective AC noise is significantly reduced compared to TARGET 5 over the whole amplitude range.

As discussed in Section \ref{sec:time_base}, most effective AC noise is due to the assumed sampling cell bin width of 1\,ns. The residuals' non-linear behavior suggests an additional noise component, which was not observable in TARGET 5, possibly due to the larger noise level. A possible way to improve the effective AC noise is presented in the following section.

\subsubsection{AC Correction transfer function}
\label{sec:AC_corr}

%To tackle the AC noise while measuring shaped SiPM pulses up to 80\,MHz
One way to improve the AC noise for pulse-shaped data that emerge in SiPM measurements is an AC Correction transfer function. The objective here is to directly correct the observable, such as the amplitude or the integral of the pulse. The advantage is that it reduces the noise and is sensitive to possible bias and gain mismatches from internal and external components. An external AC correction is essential as the absolute accuracy is not measurable with an internal DC calibration if for example a shaping circuit is used in front of the ASICs. 

As a function generator, the Keysight 33622A is used, generating Gaussian-shaped pulses of 10\,ns width similar to the expected output of the shaped photo multiplier pulses. With the signal splitter board, the pulses are injected parallel to the 16 channels of the evaluation board. The external trigger is used to cover all storage cells equally in all of their digitization phases.

To extract the pulses digitized by CTC, the pulses are upsampled by a factor of ten to allow a finer adjustment of the applied matched filter (Gaussian). The maximum amplitude of the filtered signal is assigned to the storage cell in which it occurred. The positive dynamic range of 1.65\,V appears smaller because the matched filter artificially shrinks the amplitude. Around 100,000 waveforms are required (corresponding to around 25 maxima per storage cell), to obtain a mean standard error of 0.2\,mV in a range of \mbox{0 - 1.3\,V.} In the saturation regime, the error increases up to 0.8\,mV until the hard cap saturation due to the limited positive dynamic range of 1.65\,V of the Wilkinson ADC reduces it again.

\begin{figure}[t]
\centering
\includegraphics[width=0.48\textwidth]{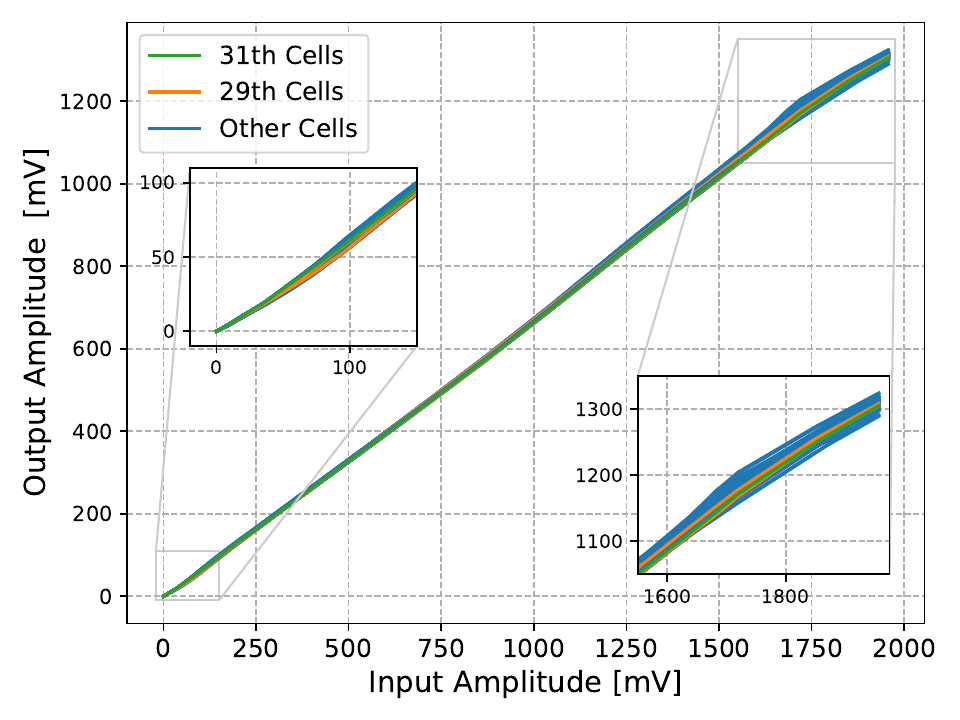}
\caption{\label{fig:AC_CORR_example} Example of AC correction transfer functions of SN0001 with both non-linear regimes highlighted. A difference in 29th and 31st cell behavior is not observed.}
\end{figure}

An example AC correction transfer function of SN0001 of channel 0 and all storage cells is shown in Figure \ref{fig:AC_CORR_example}. While linear from 50\,mV to 2\,V, it saturates smoothly due to the Gaussian filter. The AC correction transfer function is only valid if the pulses are extracted in the same manner. To not dictate the charge extraction method, it is possible to scale the entire waveform by the AC correction to address any possible bias without worsening the signal-to-noise ratio.

\begin{figure}[t]
\centering
\includegraphics[width=0.48\textwidth]{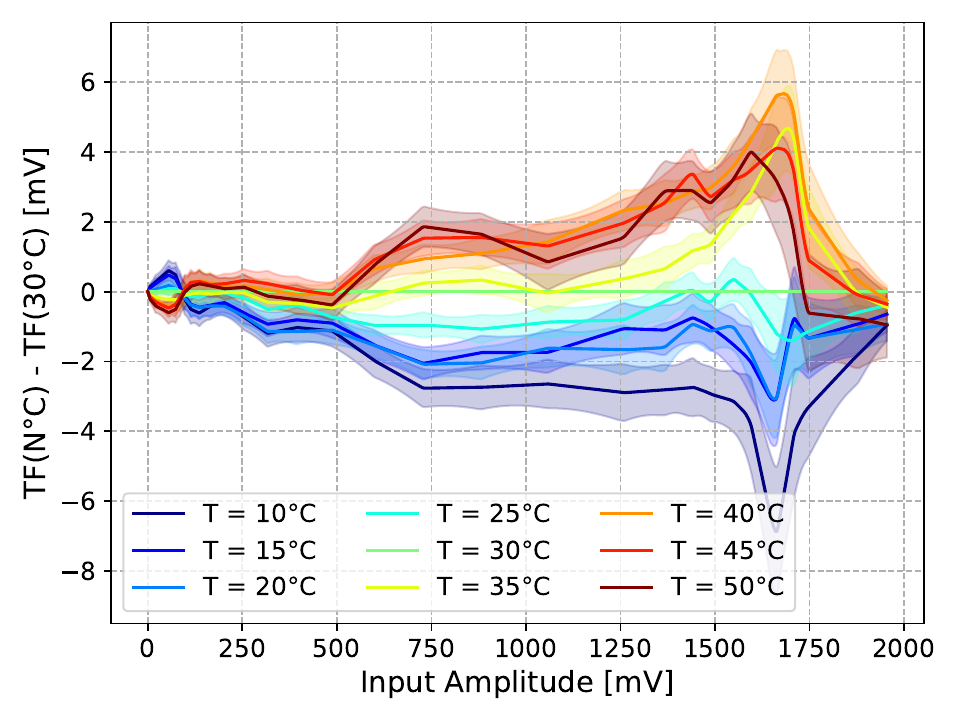}
\caption{\label{fig:AC_CORR_T_dep} The temperature dependence of the AC correction transfer function is given as the voltage difference between the transfer functions at different temperatures of SN0002. The standard deviation of all channels is given as a semi-transparent band.}
\end{figure}

As the AC correction relies on external pulses, re-calibration while operated in a camera is not possible. Therefore the AC correction must be temperature stable or recorded at different temperatures in the lab prior to installation in the camera. The temperature analysis results are presented in Figure \ref{fig:AC_CORR_T_dep}. The transfer function is sufficiently stable with a temperature-induced spread below $\pm 1$\,mV for amplitudes below 500\,mV and a relative error below 0.5\,\% over four decades of degrees Celsius. The spike in the saturation regime of the transfer function at 1650\,mV emerges from the lack of measured amplitudes where the slope changes.
The evaluation of the performance under different temperatures is presented in Section \ref{sec:pulse_performance}.

\subsubsection{\label{sec:pulse_performance}Performance in pulse mode operation}

\begin{figure}[t]
\centering
\includegraphics[width=0.48\textwidth]{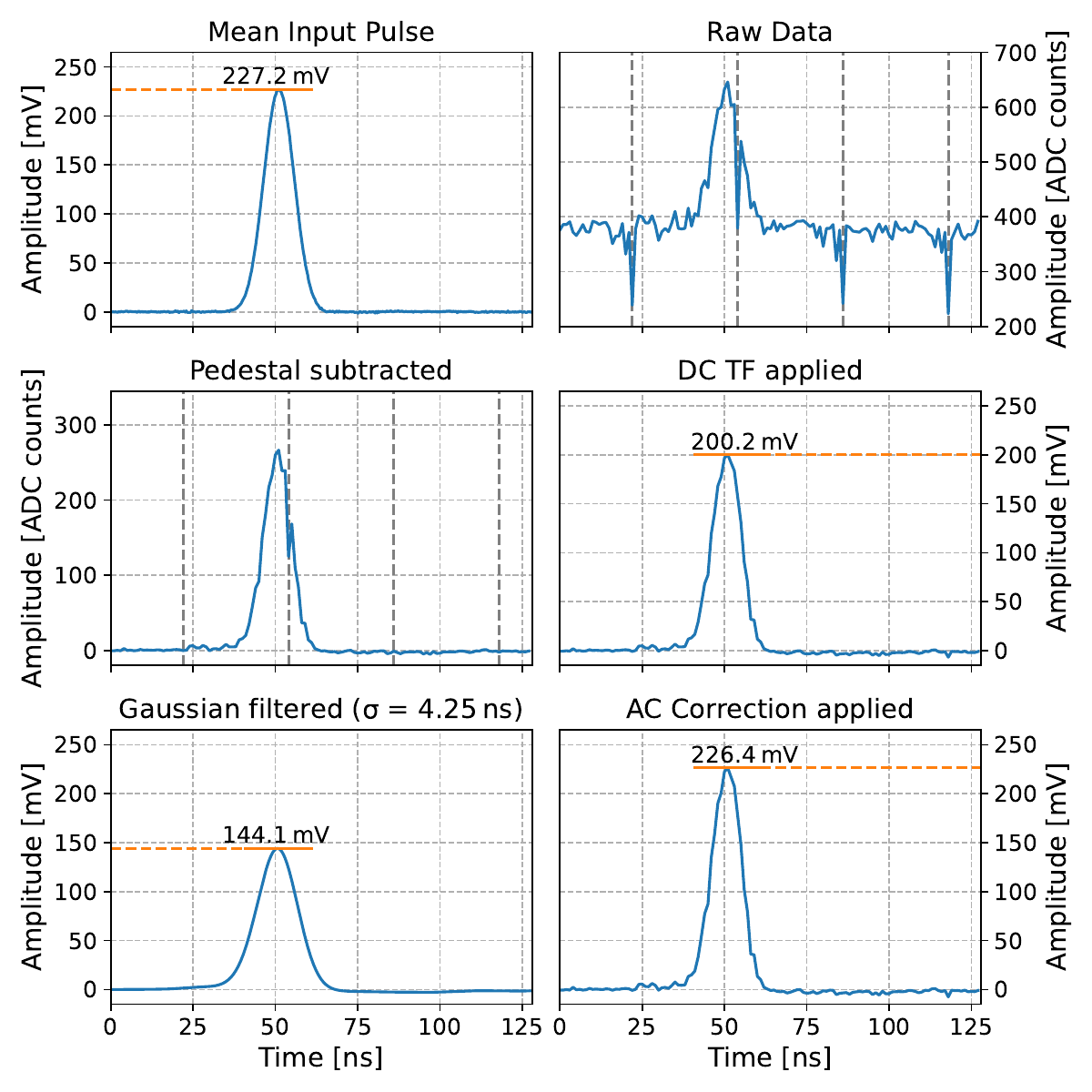}
\caption{\label{fig:2dhist}  Subsequent calibration steps applied to a waveform of a Gaussian-shaped pulse on SN0001. Top left, the mean input pulse is shown, digitized with a Tektronix MSO064b Oscilloscope. The raw digitized waveform is shown on the top right. Baseline subtraction is shown on the center left. Center right shows the waveform after the DC transfer function is applied. On the bottom left, a Gaussian filter with 1.5\,ns width is applied. The input pulse amplitude was 227.2\,mV and the amplitude of 200.2\,mV is corrected to 226.4\,mV by the AC correction as seen on the panel bottom right.}
\end{figure}

To evaluate the performance of the CTC under realistic settings, it is tested in pulse mode operation with  Gaussian-shaped pulses of 10\,ns (FWHM) width. For this, data sets at different temperatures were recorded with amplitudes explicitly different from the ones used in the AC correction, probing the worst possible outcome. An example waveform of input amplitude 227.2\,mV is shown in Figure \ref{fig:2dhist} with different calibration steps applied. In the raw waveform, the pulse is visible with digitization artifacts. The structure of the 31st cell, seen as negative spikes, dominates the picture. After operational pedestal subtraction, the ASIC cell-to-cell variations in the baseline are largely removed, leading to a visibly cleaner waveform and improved signal-to-noise. However, the scaling in ADC is still non-linear, and the 31st cell of each block shows significantly worse behavior than the others. The block-dependent DC transfer function fixes both but leaves a bias of 27.0\,mV. Applying the AC correction to the amplitude corrects it to 226.4\,mV, matching the input amplitude to 0.3\,\%.

\begin{figure}[t]
\centering
\includegraphics[width=0.48\textwidth]{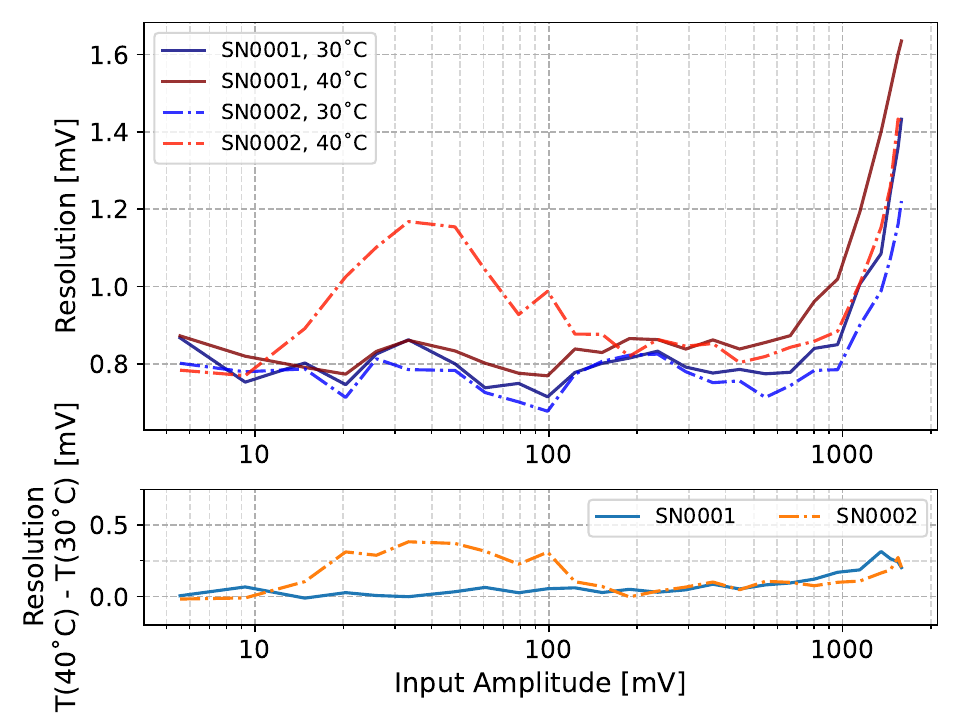}
\caption{\label{fig:performanceresolution} Fractional resolution for both fully calibrated modules at two different temperatures (with the exception of the AC correction, which is fixed at 30\,$^\circ$C).}
\end{figure}

The resolution as a function of amplitude for fully calibrated ASICs can be seen in Figure \ref{fig:performanceresolution} for both evaluation boards and two temperatures, with the temperature-induced difference in the bottom panel. For the 40\,$^\circ$C runs, an AC correction transfer function taken at 30\,$^\circ$C is used to test the temperature stability of the proposed calibration technique. 

SN0001 shows a near constant resolution of $\leq$ 1\,mV from 5\,mV to 1000\,mV for both temperatures. Only for amplitudes above 1\,V, where the slope of the DC transfer functions increases, the resolution worsens linearly up to 1.6\,mV towards the end of the input voltage range. Saturated amplitudes could in principle be extracted, but this goes beyond the scope of this paper.

For SN0002, the resolution looks similar to that of SN0001 for the 30\,$^\circ$C data set. However, at 40\,$^\circ$C, the resolutions worsen up to 1.2\,mV between 10\,mV and 100\,mV. This is to be expected, as the tuning of the ASIC parameters was done on SN0001 and applied to SN0002. Manufacturing tolerances, in general, and slightly mismatched parameters, lead to a less robust calibration. Further modules must be tested to characterize these ASIC to ASIC variations, but SN0002 is a precedent for all the ASICs to follow.

\begin{figure}[t]
\centering
\includegraphics[width=0.48\textwidth]{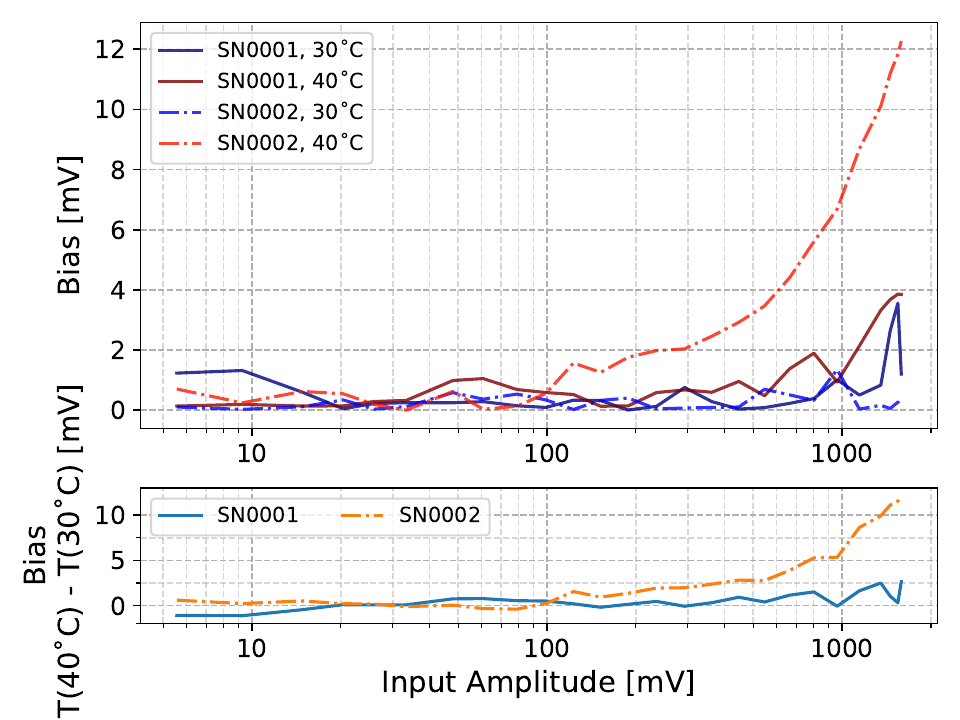}
\caption{\label{fig:performanceoffset}  Bias for both fully calibrated modules at two different temperatures (with the exception of the AC correction, which is fixed at 30\,$^\circ$C). The bias is defined as the absolute difference between the measured amplitude and the true amplitude.}
\end{figure}

Figure \ref{fig:performanceoffset} shows the reconstruction quality achieved with the calibrated module. The bias is defined here as the absolute difference between the measured amplitude and the true amplitude. Similar to the resolution of the SN0001, the bias stays below 1\,mV for nearly two magnitudes of input amplitudes for both temperatures, with slightly higher values below 10\,mV. Above 600\,mV the bias grows again up to 4\,mV at 1.6\,V input amplitude. The temperature induces difference in reconstruction quality swings between 3\,mV for the whole voltage range of SN0001.

SN0002 performs even better at 30\,$^\circ$C, where the data matches the calibration data, and reconstructs with an accuracy of $\leq$ 1\,mV over the whole input voltage range. However, it lacks temperature stability, as the bias increases linearly from 1\,mV at 100\,mV input amplitude to 12\,mV towards the end of the input voltage range. The absolute numbers look more daunting than they actually are, as the relative bias never crosses the \% mark above 100\,mV input amplitude. Again, more ASIC-to-ASIC statistics are needed to better understand the SN0002 behavior at 40\,$^\circ$C.

\begin{figure}[t]
\centering
\includegraphics[width=0.48\textwidth]{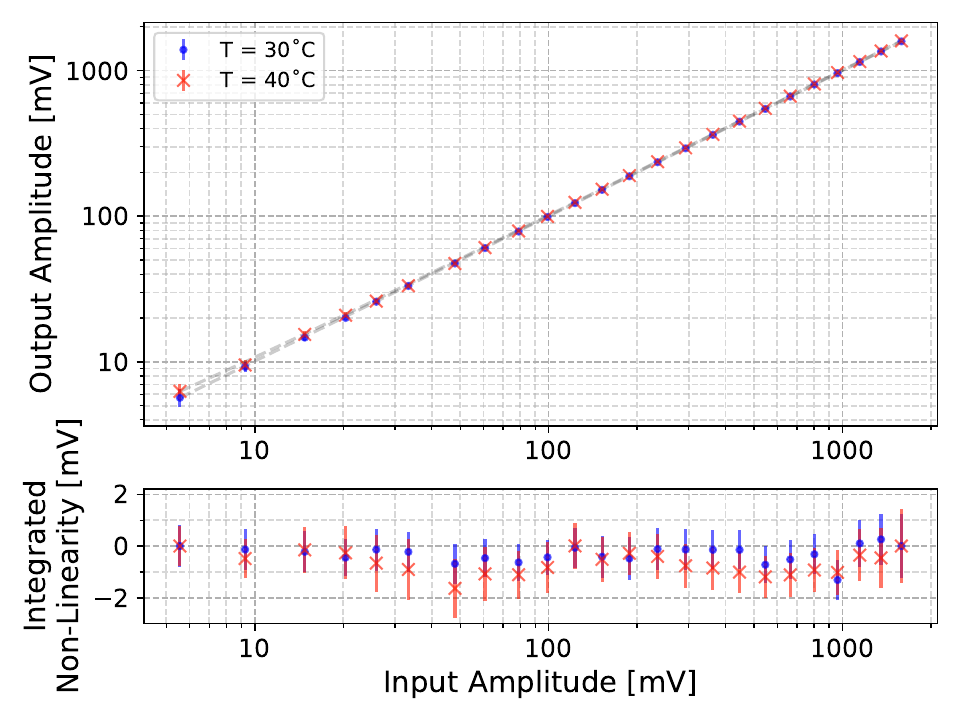}
\caption{\label{fig:InOut} Response of the fully calibrated SN0001 to different input amplitudes of 10\,ns Gaussian-shaped pulses. The integrated non-linearity is given at the bottom with the channel-to-channel spread as error bars.}
\end{figure}

The measured amplitude as a function of the input amplitude of SN0002 after full calibration is presented in Figure \ref{fig:InOut} for two different temperatures. The integrated non-linearity is measured by subtracting each measurement point from a linear function connected by the first and last point. It is shown in the bottom panel of Figure \ref{fig:InOut} with the standard deviation of the channel-to-channel spread shown as error bars. Although the response of the ASIC changes, the linearity is given for the effective dynamic range and both temperatures. It only exceeds the 1\,mV mark at around 1\,V and shows a mild deviation for a temperature mismatch of 10\,$^\circ$C in the AC correction transfer function.

All in all, the calibration procedure is deemed temperature stable and the obtained performance results indicate that a single AC correction transfer function at a moderate temperature is satisfactory. The response is linear independent of the measured temperatures. The temperature-induced change in resolution at ten degree difference is 0.1\,mV averaged over all input amplitudes and approximately worsens with rising amplitude. Only the bump in SN0002 shows an anomaly of 0.4\,mV between 10\,mV and 100\,mV. The reconstruction quality worsens around 1.5\,mV on average between both temperatures, with again SN0002 as an outlier. Above 100\,mV, it grows up to 12\,mV additional bias, although it never exceeds the 1\,\% mark in relative numbers.

\subsubsection{\label{sec:IACT}Suitability for IACT Cameras}

To evaluate the performance in the context of an IACT camera the intensity resolution for the SST camera defined by the CTAO requirement \citep{Jama} is used as a comparison. It is defined as the following

\begin{align*}
    \frac{\sigma_\mathrm{I}}{\mathrm{I_\mathrm{T}}} &= \frac{1}{\mathrm{I_\mathrm{T}}} 
                                                        \sqrt{\frac{   \sum_{i=0}^{\mathrm{N}  }(\mathrm{I}_{\mathrm{M}_i} - \mathrm{I}_{\mathrm{T}})^2}{N}},
\end{align*}
where $\frac{\sigma_\mathrm{I}}{\mathrm{I_\mathrm{T}}}$ is the fractional intensity resolution, $\mathrm{I_\mathrm{T}}$ is the true intensity - the true number of photons impacting the camera per event, $\mathrm{I}_{\mathrm{M}_i}$ the measured intensity of the event and N the number of measured events. As only the digitizer in the form of CTC is characterized here, a quantum efficiency of 40\,\% for the Silicon photomultipliers combined with the window transmission and a scaling of 3.37\,mV per photo electron is assumed, representing the target values for the SST camera. Therefore, this process translates an intensity of one photon to an amplitude of roughly 1.35\,mV. In addition to the intensity resolution of CTC, the Poisson error must be quadratic. The data set from Section \ref{sec:pulse_performance} is used for comparison.

\begin{figure}[t]
\centering
\includegraphics[width=0.48\textwidth]{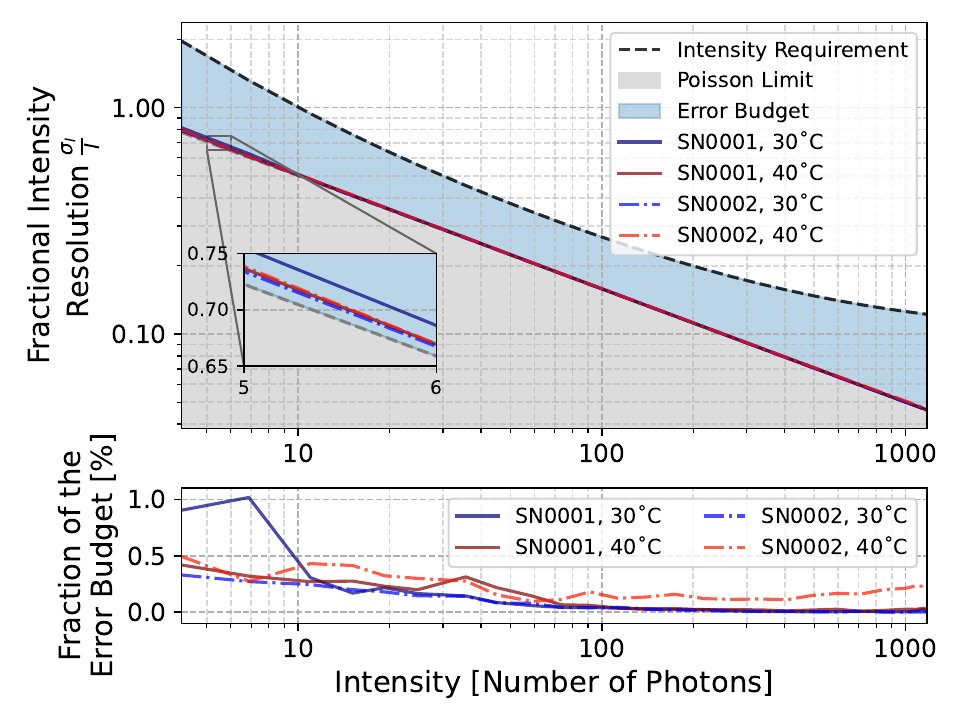}
\caption{\label{fig:IntRes} Intensity resolution of fully calibrated modules with the SST requirement in black. The Poisson Limit is marked as the gray area. The performance of the whole SST camera must be in the blue area, defined as error budget as multiple components contribute to it. A quantum efficiency of 40\,\% for the photomultipliers is assumed with a 3.37\,mV pulse per photo electron conversion.}
\end{figure}

The results for SN0001 and SN0002 are shown in Figure \ref{fig:IntRes}. The minimal intensity resolution is defined by the Poisson error; intensity resolutions in the gray area are therefore not possible. On top as a black dashed line is the SST requirement. The blue area between them is the error budget, which can be calculated by quadratically subtracting the Poisson limit from the requirement, as the contributions are independent of each other. Both modules are nearly indistinguishable from the Poisson limit.

For better depiction, the fraction of used error budget (UEB) is given in the bottom panel as 
\begin{align*}
    \mathrm{UEB} &= 1 - \sqrt{1-\frac{\sigma^2_\mathrm{I}}{\sigma^2_\mathrm{Requirement} - \sigma^2_\mathrm{Poisson}}}.
\end{align*}
This is derived by dividing the quadratic difference between the requirement and the fractional intensity resolution by the error budget. The contribution of CTC is below 0.5\,\% over the whole span of given photon numbers for both modules and temperature data sets, except for SN0001 at 30\,$^\circ$C under ten photons. The temperature mismatch worsens the performance, but not to a significant level as the contribution from CTC itself is not significant, although the in Section \ref{sec:pulse_performance} discussed performance losses due to temperature mismatch in SN0002 can clearly be seen. Between ten and 100 photons, the decline in resolution is the leading factor for the worse performance, while for photon numbers above 100 it is the additional bias.
Nevertheless, the TARGET ASICs are the least contributing noise factor in the signal chain of the SST. Additional contributions to the error budget will be the night sky background (NSB), photodetector noise, front-end electronics uncertainties, optical cross-talk and unaccounted temperature dependencies of different components.

Concluding, CTC is perfectly suited to fulfil the given CTA requirements and similar setups.

\subsection{Comparison to previous versions}

An overview of the characteristics and performance improvements compared to TARGET 5 is given in \mbox{Table \ref{tab:TARGET_table}.} From TARGET 5 to 7  the Wilkinson clock was changed to be fed externally from the FPGA, improving the temperature stability and time synchronization. The TARGET C kept the digitization part of TARGET 7, while the T5TEA is based on the TARGET 5 trigger part. The split into a sampling/digitization ASIC and a companion trigger ASIC in TARGET C combined with a modified ASIC building block architecture results in an improved trigger performance and a larger dynamic range \citep{TC_paper, TARGET_proceeding}.  Another modification of the ASIC was needed as the applied scalable CMOS design rules had to be updated to the TSMC native rules due to a manufacturer change.
Additionally, the Gray code counter of Wilkinson ADC was extended to full 12 bits\footnote{In previous versions, the Wilkinson clock itself formed the 12th bit} and a sum 16 channel trigger was implemented into CT5TEA. The digitization clock speed was reduced to 208\,MHz to avoid a possible race condition in the Gray code counter. The main performance improvement of CTC to TARGET C was done by dedicated parameter tuning, leading to an improved resolution and linearity.

\begin{table*}[t]
    \caption{ Input used from \citep{TC_paper}. For comparability to the older generation of TARGET, a scaling of 3.37\,$\frac{\mathrm{mV}}{\mathrm{p.e.}}$ is used.}
    \label{tab:TARGET_table}
    \begin{center}
    \begin{tabular}{llll}
        \hline

         & TARGET 5 & CTC + CT5TEA\\   
         \hline
        \textbf{Characteristics} & & \\
        \hline
        Number of channels & 16 & 16\\
        Sampling frequency (GSa/s) & 0.4 - 1  &  1\\
        Size of storage array & 16384 & 16384\\
        Digitization clock speed (MHz) & 700  & 208 \\
        Samples digitized simultaneously & 32 $\times$ 16 &  32 $\times$ 16\\
        Trigger (sum of four channels) & integrated & companion\\
        Power consumption per channel (mW) & 20 & $\sim$ 60 \\
        \hline
        \textbf{Performance} &  &  \\
        \hline
        Dynamic range (V) & 1.1 &  $\geq$ 2.2\\
        Integrated non-linearity (mV)  & 70 &  $\leq$ 1.6\\
        Resolution at 10 p.e. (\%) & 8 &  2.5\\
        Resolution at $>$ 100 p.e. (\%) & 2 & $\leq$ 0.3\\
        % Bandwidth [MHz] & 2 & $\leq$0.8 & $\leq$0.8 &  $\leq$0.3\\
        % Crosstalk over full Bandwidth [\%] & 2 & $\leq$0.8 & $\leq$0.8 &  $\leq$0.8\\
        Minimum trigger threshold (mV) & 20 & $\leq$ 2.5 \\
        Minimum trigger noise (mV) & 5 & $\leq$ 0.5\\
        \hline
    \end{tabular}
        \end{center}
\end{table*}

\section{Front-end electronics modules}
\label{sec:FEE}

A number of front-end electronics modules containing the CTC and CT5TEA ASICs were designed. The high integration of the ASICs allows the building of compact modules suitable for equipping the small focal plane of Schwarzschild-Couder-like Cherenkov telescopes. For the SST, 32 TARGET modules, hosting four ASIC pairs each, are integrated into the camera. In addition to the trigger and digitization functionality provided by the ASICs, the module contains signal shaping and photocurrent monitoring for each of the 64 channels and bias supply for SiPMs. The SST camera will have a field of view of 8.8$^\circ$ with 2048 individual SiPM pixels at a size of 6\,mm x 6\,mm \citep{SPIE_sst_paper}. A prototype with eight modules is currently being built and tested.

The camera of the SCT follows a similar design and aims for an even higher pixel count of 11,328 \citep{psc}. The current prototype using both TARGET 7 and TARGET C ASICs has successfully detected the Crab nebula and aims for a camera upgrade using the CTC architecture \citep{sct_proceeding}.

IceAct, a cost-effective IACT consisting of a Fresnel lens focusing the Cherenkov light onto 61 SiPMs, utilizes a modified TARGET C module. The sensitivity for hybrid cosmic ray measurements with IceCube and IceTop has been demonstrated at the south pole \citep{IceAct1, IceAct2}.

TARGET ASICs are not limited to IACTs and are also deployed in the Antarctic Demonstrator for Advanced Particle-astrophysics (ADAPT), and the proof-of-concept of the Advanced Particle-astrophysics Telescope (APT), which will run the CT5TEA ASIC and low power for space missions optimized ALPHA ASIC for sampling and digitization. 

\section{Conclusion and outlook}
\label{sec:conclusion}

The main advantage of the latest generation of TARGET ASIC is the split between the trigger and sampling/digitization path, the in-depth tuning of the control parameters, and the development of new calibration methods. Together, they are responsible for the improved performance of the TARGET ASICs. The dynamic range is increased to 2.2\,V, equivalent to an effective bit range of 11.7\,bits and 11.6\,bits for SN0001 and SN0002, respectively. With an operational pedestal voltage of 750\,mV, an effective dynamic range of 1.7\,V for pulses is possible at a baseline noise of 0.7\,mV, although baseline noise beyond 0.6\,mV is achievable with decreased positive dynamic range. Due to the new block-dependent DC transfer function, it is also possible to lower the 0.2$\frac{\mathrm{ADC}}{\mathrm{K}}$ drift while the ASIC is operating.

The - 3\,dB bandwidth of 220\,MHz is sufficient for pulse operation down to 3.6\,ns width. The cross-talk is reduced to $\leq$ 1\,\% over the whole - 3\,dB bandwidth down to 0.2\,\% for large amplitude pulses. It is mainly observed in the next-neighbor channel due to the layout of the evaluation board, not the ASIC itself.

With the new AC correction, the resolution improves to 2.5\,\% at 10\,p.e. and $\leq$ 0.2\,\% at 100\,p.e. at a conversion factor of 3.37 $\frac{\mathrm{mV}}{\mathrm{p.e.}}$ and is temperature-stable over several degrees with an integrated non-linearity of 1.6\,mV. This is an order of magnitude improved compared to previous generations of the ASIC.

The trigger performance has also improved by an order of magnitude compared to the TARGET 5 due to the separation into two separate ASICs. The minimum achievable effective trigger threshold is now \mbox{(2.80\,mV $\pm$ 0.15)\,mV,} significantly below one p.e.. The trigger noise for a 33.7\,mV (10\,p.e.) effective trigger threshold is down to \mbox{(0.52\,mV $\pm$ 0.16)\,mV}.%, less than $\frac{\mathrm{1}}{\mathrm{6}}$ of a p.e..

Within the CTAO project, the CTC/CT5TEA ASICs are ready to be deployed in the SST and SCT cameras.

\section*{Acknowledgement}
We gratefully acknowledge financial support from the agencies and organizations listed at \url{https://www.ctao.org/for-scientists/library/acknowledgments/}.

\addcontentsline{toc}{section}{References}
\markboth{Bibliography}{Bibliography}
\bibliography{references.bib}
%\end{appendices}


% \begin{thebibliography}{99}

% \bibitem{a}
% Author, \emph{Title}, \emph{J. Abbrev.} {\bf vol} (year) pg.

% \bibitem{b}
% Author, \emph{Title},
% arxiv:1234.5678.

% \bibitem{c}
% Author, \emph{Title},
% Publisher (year).

% Please avoid comments such as "For a review'', "For some examples",
% "and references therein" or move them in the text. In general,
% please leave only references in the bibliography and move all
% accessory text in footnotes.

% Also, please have only one work for each \bibitem.

% \end{thebibliography}
\end{document}